\documentclass[prc,twocolumn,twoside,showpacs]{revtex4}

\usepackage{graphicx}
\usepackage{amsmath,amssymb,bm,color}
\usepackage{times}

\newcommand{\eq}[1]{\begin{equation}#1\end{equation}}
\newcommand{\eqmulti}[1]{\begin{equation}\begin{split}#1\end{split}\end{equation}}

\newcommand{\bra}[1]{\ensuremath{\langle{#1}|\,}}
\newcommand{\ket}[1]{\ensuremath{\,|{#1}\rangle}}
\newcommand{\braket}[2]{\ensuremath{\langle{#1}|{#2}\rangle}}

\newcommand{\matrixe}[3]{\ensuremath{\langle{#1}|\,{#2}\,|{#3}\rangle}}

\newcommand{\comm}[2]{\ensuremath{[{#1},{#2}]}}

\newcommand{\op}[1]{\ensuremath{\mathrm{#1}}}
\newcommand{\adj}[1]{\ensuremath{{{#1}}^{\dag}}}
\newcommand{\corr}[1]{\ensuremath{\widetilde{#1}}}

\newcommand{\ii}{\ensuremath{\mathrm{i}}}
\newcommand{\dd}{\ensuremath{\mathrm{d}}}

\renewcommand{\vec}[1]{\ensuremath{\bm{#1}}}

\newcommand{\cO}{\ensuremath{\op{c}}}
\newcommand{\ccO}{\ensuremath{\adj{\op{c}}}}
\newcommand{\gO}{\ensuremath{\op{g}}}
\newcommand{\hO}{\ensuremath{\op{h}}}

\newcommand{\qO}{\ensuremath{\op{q}}}
\newcommand{\rO}{\ensuremath{\op{r}}}
\newcommand{\tO}{\ensuremath{\op{t}}}
\newcommand{\vO}{\ensuremath{\op{v}}}

\newcommand{\AO}{\ensuremath{\op{A}}}
\newcommand{\CO}{\ensuremath{\op{C}}}
\newcommand{\CCO}{\ensuremath{\adj{\op{C}}}}

\newcommand{\OO}{\ensuremath{\op{O}}}

\newcommand{\TO}{\ensuremath{\op{T}}}
\newcommand{\VO}{\ensuremath{\op{V}}}

\newcommand{\PiO}{\ensuremath{\op{\Pi}}}

\newcommand{\NC}{\ensuremath{\mathcal{N}}}

\newcommand{\aOV}{\ensuremath{\vec{\op{a}}}}
\newcommand{\bOV}{\ensuremath{\vec{\op{b}}}}

\newcommand{\pOV}{\ensuremath{\vec{\op{p}}}}
\newcommand{\qOV}{\ensuremath{\vec{\op{q}}}}
\newcommand{\rOV}{\ensuremath{\vec{\op{r}}}}

\newcommand{\xOV}{\ensuremath{\vec{\op{x}}}}

\newcommand{\LOV}{\ensuremath{\vec{\op{L}}}}

\newcommand{\SOV}{\ensuremath{\vec{\op{S}}}}

\newcommand{\sigmaOV}{\ensuremath{\vec{\op{\sigma}}}}

\newcommand{\tautauO}{\ensuremath{(\vec{\op{\tau}}_{\!1}\!\cdot\!\vec{\op{\tau}}_{\!2})}}
\newcommand{\sigmasigmaO}{\ensuremath{(\vec{\op{\sigma}}_{\!1}\!\cdot\!\vec{\op{\sigma}}_{\!2})}}
\newcommand{\tensorO}{\ensuremath{\op{s}_{12}}}
\newcommand{\tensorRRO}{\ensuremath{\op{s}_{12}(\tfrac{\rOV}{\rO},\tfrac{\rOV}{\rO})}}
\newcommand{\tensorLLO}{\ensuremath{\op{s}_{12}(\LOV,\LOV)}}
\newcommand{\tensorQQO}{\ensuremath{\op{s}_{12}(\qOV_{\Omega},\qOV_{\Omega})}}
\newcommand{\tensorRQO}{\ensuremath{\op{s}_{12}(\rOV,\qOV_{\Omega})}}
\newcommand{\tensorbarQQO}{\ensuremath{\bar{\op{s}}_{12}(\qOV_{\Omega},\qOV_{\Omega})}}
\newcommand{\spinorbitO}{\ensuremath{(\vec{\op{L}}\cdot\vec{\op{S}})}}
\newcommand{\orbitsqrO}{\ensuremath{\vec{\op{L}}^2}}

\newcommand{\pmS}{\ensuremath{\!\pm\!}}
\newcommand{\mpS}{\ensuremath{\!\mp\!}}

\newcommand{\Rm}{\ensuremath{R_-}}
\newcommand{\DRm}{\ensuremath{R'_-}}

\newcommand{\Rp}{\ensuremath{R_+}}
\newcommand{\DRp}{\ensuremath{R'_+}}
\newcommand{\DDRp}{\ensuremath{R''_+}}
\newcommand{\DDDRp}{\ensuremath{R'''_+}}

\newcommand{\Rpm}{\ensuremath{R_{\pm}}}
\newcommand{\Rmp}{\ensuremath{R_{\mp}}}

\newcommand{\RRp}{\ensuremath{\mathcal{R}_+}}

\newcommand{\UCOM}{\ensuremath{\textrm{UCOM}}}
\newcommand{\elem}[2]{\ensuremath{{}^{#2}\text{#1}}}

\definecolor{FGViolet}{rgb}{0.61,0.32,0.61}
\definecolor{FGBlue}{rgb}{0,0,0.8}
\definecolor{FGGreen}{rgb}{0.2,0.7,0.2}
\definecolor{FGRed}{rgb}{0.8,0,0}
\definecolor{FGLightGray}{rgb}{0.85,0.85,0.85}
\definecolor{FGGray}{rgb}{0.6,0.6,0.6}

\newcommand{\linemediumsolid}[1][black]{\unitlength1ex
  ({\color{#1}\begin{picture}(6,1)
  \linethickness{0.4mm}
  \put(0,0.5){\line(1,0){6.0}}
  \end{picture}})\nolinebreak
}

\newcommand{\linemediumdashed}[1][black]{\unitlength1ex
  ({\color{#1}\begin{picture}(6,1)
  \linethickness{0.4mm}
  \put(0,0.5){\line(1,0){1.5}}
  \put(2.2,0.5){\line(1,0){1.5}}
  \put(4.4,0.5){\line(1,0){1.5}}
  \end{picture}})\nolinebreak
}

\newcommand{\linemediumdotted}[1][black]{\unitlength1ex
  ({\color{#1}\begin{picture}(6,1)
  \linethickness{0.4mm}
  \put(0,0.5){\line(1,0){0.4}}
  \put(0.9,0.5){\line(1,0){0.4}}
  \put(1.8,0.5){\line(1,0){0.4}}
  \put(2.7,0.5){\line(1,0){0.4}}
  \put(3.6,0.5){\line(1,0){0.4}}
  \put(4.5,0.5){\line(1,0){0.4}}
  \put(5.4,0.5){\line(1,0){0.4}}
  \end{picture}})\nolinebreak
}

\newcommand{\linemediumdashdot}[1][black]{\unitlength1ex 
  ({\color{#1}\begin{picture}(6,1)
  \linethickness{0.4mm}
  \put(0,0.5){\line(1,0){0.4}}
  \put(0.9,0.5){\line(1,0){1.5}}
  \put(2.9,0.5){\line(1,0){0.4}}
  \put(3.8,0.5){\line(1,0){1.5}}
  \put(5.8,0.5){\line(1,0){0.4}}
  \end{picture}})\nolinebreak
}

\begin{document}

\title{Matrix Elements and Few-Body Calculations within the Unitary Correlation Operator Method}

\author{R. Roth}
\email{robert.roth@physik.tu-darmstadt.de}
\author{H. Hergert}
\author{P. Papakonstantinou} 

\affiliation{Institut f\"ur Kernphysik, Technische Universit\"at Darmstadt,
64289 Darmstadt, Germany}

\author{T. Neff}
\author{H. Feldmeier} 

\affiliation{Gesellschaft f\"ur Schwerionenforschung, Planckstr. 1, 
64291 Darmstadt, Germany}

\date{\today}

\begin{abstract}    
We employ the Unitary Correlation Operator Method (UCOM) to construct
correlated, low-momentum matrix elements of realistic nucleon-nucleon
interactions. The dominant short-range central and tensor correlations
induced by the interaction are included explicitly by an unitary
transformation. Using correlated momentum-space matrix elements of the
Argonne V18 potential, we show that the unitary transformation eliminates
the strong off-diagonal contributions caused by the short-range repulsion
and the tensor interaction, and leaves a correlated interaction dominated
by low-momentum contributions.  We use correlated harmonic oscillator
matrix elements as input for no-core shell model calculations for
few-nucleon systems. Compared to the bare interaction, the convergence
properties are dramatically improved. The bulk of the binding energy can
already be obtained in very small model spaces or even with a single
Slater determinant. Residual long-range correlations, not treated
explicitly by the  unitary transformation, can easily be described in
model spaces of moderate size allowing for fast convergence.  By
varying the range of the tensor correlator we are able to map out the Tjon
line and can in turn constrain the optimal correlator ranges.
\end{abstract}

\pacs{21.30.Fe, 21.60.-n, 13.75.Cs}

\maketitle

\clearpage
\section{Introduction}
\label{sec:intro}

One of the prime challenges in modern nuclear structure theory is the
description of properties of nuclei across the whole nuclear chart based
on realistic nucleon-nucleon interactions. Several modern nucleon-nucleon 
interactions that reproduce the experimental two-body data with high 
precision are available, e.g., the Argonne V18 potential \cite{WiSt95}, 
the CD Bonn potential \cite{Mach01}, or the Nijmegen potentials
\cite{StKl94}. The use of these interactions for nuclear
structure calculations in a strict \emph{ab initio} fashion is restricted
to light isotopes, where Green's Function Monte Carlo
\cite{PiWi04,PiVa02,PiWi01} or no-core shell model calculations
\cite{NaOr02,CaNa02,NaVa00b} are computationally feasible.  These
virtually exact solutions of the nuclear many-body problem show that
realistic NN-potentials supplemented by a phenomenological three-nucleon
force are able to reproduce experimental ground states and excitation spectra
of light nuclei. Furthermore, recent developments in chiral perturbation theory 
provide schemes to construct two- and three-nucleon forces systematically
\cite{EnMa03,EpNo02}.  

A major obstacle for \emph{ab initio} nuclear structure calculations are
the strong  short-range correlations induced by realistic NN-interactions.
There are two dominant components: (\emph{i}) correlations induced by the
short-range repulsive core in the central part of the potential and
(\emph{ii}) correlations generated by the strong tensor interaction. It is
well known that, in a shell-model language, the description of these
correlations requires extremely large model-spaces --- the repulsive core
and the tensor interaction lead to sizable admixtures of high-lying
shells. Simple many-body spaces, which remain tractable for large particle
numbers, cannot describe these correlations. In the extreme case, e.g. in
a Hartree-Fock approach, the many-body state is restricted to a single
Slater determinant which is not capable of representing these
correlations by construction. Therefore, the use of a bare realistic
NN-interaction in such a framework has to fail. 

There are several recent attempts to tackle this problem. One is the
so-called $V_{\text{low}k}$ approach \cite{BoKu03,BoKu03b}, which employs
renormalization group techniques to reduce the bare realistic potential to
a low-momentum interaction. Effectively, the high-momentum contributions, 
which are responsible for the admixture of high-lying states, are
integrated out leaving an effective low-momentum interaction suitable for
small model spaces. 

Another approach is the Unitary Correlation Operator Method (UCOM)
\cite{RoNe04,NeFe03,FeNe98}. Here the short-range central and tensor
correlations are explicitly described by a state and
basis-independent unitary transformation. Applying the unitary
operator of the transformation to uncorrelated many-body states, e.g., the
Slater determinant of a Hartree-Fock scheme, leads to a new correlated
state which has the dominant short-range correlations built in.
Alternatively, the correlation operator can be applied to the Hamiltonian,
leading to a phase-shift equivalent correlated interaction $\VO_{\UCOM}$ which is
well suited for small low-momentum model spaces. Hence it can be used as an
universal input for a variety of many-body methods. The operator form of this
correlated interaction resulting for the Argonne V18 (AV18) potential has
been used successfully to perform nuclear structure calculations in the
framework of Fermionic Molecular Dynamics \cite{RoNe04,NeFe04,NeFe05}. 

In this paper we are going to apply the Unitary Correlation Operator
Method to derive correlated two-body matrix elements. They serve as convenient and
universal input for a variety of many-body techniques, ranging from Hartree-Fock
to shell-model. Following a summary of the formalism of the
Unitary Correlation Operator Method in Sec. \ref{sec:ucom}, we derive
explicit expressions for correlated matrix elements in Sec.
\ref{sec:corrme}. Optimal correlation functions for the AV18 potential are
constructed in Sec. \ref{sec:optcorr} and the properties of the correlated
momentum-space matrix elements are discussed in Sec. \ref{sec:momentumme}.
Finally, in Sec. \ref{sec:ncsm}, we present results of no-core shell model
calculations using correlated oscillator matrix elements, which highlight
the effect of the unitary transformation and the properties of the
correlated interaction.

\section{The Unitary Correlation Operator Method (UCOM)}
\label{sec:ucom}

\subsection{Unitary Correlation Operator}
\label{sec:ucomop}

The concept of the Unitary Correlation Operator Method
\cite{RoNe04,NeFe03,FeNe98} can be summarized as follows: The dominant 
short-range central and tensor correlations are  imprinted into a simple
many-body state $\ket{\Psi}$ through a state-independent unitary
transformation 
\eq{ \label{eq:corr_state}
  \ket{\corr{\Psi}} = \CO\; \ket{\Psi} \;.
}
The unitary correlation operator $\CO$ describing this transformation is
given in an explicit operator form, independent of the particular
representation or model space. The correlated many-body state explicitly
contains the important short-range correlations generated by the
interaction. Even if we start with a simple Slater determinant as
uncorrelated state  $\ket{\Psi}$ then the correlated state
$\ket{\corr{\Psi}}$ cannot be represented by a single or a superposition
of few Slater determinants anymore. 

When calculating expectation values or matrix elements of some operator
$\AO$ using correlated states 
\eq{ \label{eq:corr_matrixelem}
  \matrixe{\corr{\Psi}}{\AO}{\corr{\Psi}'}
  = \matrixe{\Psi}{\CCO\AO\CO}{\Psi'}
  = \matrixe{\Psi}{\corr{\AO}}{\Psi'} \;,
}
we can define a correlated operator through the similarity transformation
\eq{ \label{eq:corr_operator}
  \corr{\AO} 
  = \CO^{-1}\AO\,\CO 
  = \CCO\AO\,\CO \;.
}
Due to the unitarity of $\CO$ the notions of correlated states and
correlated operators are equivalent and we may choose the form that is
technically more advantageous.

In the case of the nuclear many-body problem, the unitary correlation
operator $\CO$ has to account for short-range central and tensor
correlations as outlined in Sec. \ref{sec:intro}. It is convenient to
disentangle these different types of correlations and define the
correlation operator as a product of two unitary operators,
\eq{
  \CO 
  = \CO_{\Omega} \CO_{r} \;,
}
where $\CO_{\Omega}$ describes short-range tensor correlations and
$\CO_{r}$ central correlations. Each of these unitary operators is
written as an exponential of a Hermitian two-body generator
\eq{ \label{eq:correlator}
  \CO_{\Omega} 
  =  \exp\!\Big[-\ii \sum_{i<j} \gO_{\Omega,ij} \Big] \;,\quad
  \CO_{r}
  = \exp\!\Big[-\ii \sum_{i<j} \gO_{r,ij} \Big]\;.
}
The construction of the generators $\gO_r$ and $\gO_{\Omega}$, which
encode the relevant physics of short-range interaction-induced
correlations, is crucial.

We start with the generator $\gO_{r}$ associated with the short-range
central correlations induced by the repulsive core in the central part of
the NN-interaction. At small relative distances, the two-body density is 
strongly suppressed as a result of the repulsive core. Pictorially, the
core keeps the nucleons apart from each other so that they reside at larger
distances outside the short-range repulsion \cite{RoNe04,FeNe98}. These
correlations can be imprinted into an uncorrelated many-body state by an
unitary distance-dependent shift along the relative coordinate for each
particle pair. Such radial shifts are generated by the projection of the
relative momentum $\qOV  = \frac{1}{2} [\pOV_1 - \pOV_2]$ onto the
distance vector $\rOV = \xOV_1 - \xOV_2$ of two particles:
\eq{
  \qO_r 
  = \frac{1}{2} \big[\tfrac{\rOV}{\rO}\cdot\qOV 
    + \qOV\cdot\tfrac{\rOV}{\rO} \big] \;.
}
The distance-dependence of the shift -- large shifts at small distances
within the core, small or no shifts outside the core -- is described by a
function $s_{ST}(r)$ for each spin-isospin channel. Their shape depends on
the potential under consideration and contains all information on the
short-range central correlations. The determination of the $s_{ST}(r)$ is
discussed in detail in Sec. \ref{sec:optcorr}. The full generator for 
the central correlations reads
\eq{ \label{eq:central_generator}
  \gO_r 
  = \sum_{S,T} \frac{1}{2} 
    [ s_{ST}(\rO)\, \qO_r + \qO_r\, s_{ST}(\rO) ]\, \PiO_{ST}\;,
}
where $\PiO_{ST}$ is the projection operator onto two-body spin $S$ and
isospin $T$. 

The correlations induced by the tensor part of the interaction are of a
more complicated nature. They entangle the spins of the two nucleons with
the direction of their relative distance vector $\rOV$. Depending on the
orientations of the spins, the nucleons are shifted perpendicular to the
relative distance vector \cite{RoNe04,NeFe03}. Such shifts are generated
by the residue of the relative  momentum operator after subtracting the
radial component
\eq{
  \qOV_{\Omega} 
  = \qOV - \frac{\rOV}{\rO} \qO_r
  = \frac{1}{2\rO^2}(\LOV\times\rOV - \rOV\times\LOV) \;.
}
This ``orbital momentum'', embedded into a tensor operator which encodes the
complicated entanglement between spatial and spin degrees of freedom, enters into 
the generator of the tensor correlations
\eq{ \label{eq:tensor_generator}
  \gO_{\Omega} 
  = \sum_{T} \vartheta_T(r)\; \tensorRQO\; \PiO_{1T} 
}
using the general definition
\eqmulti{
  \tensorO(\aOV,\bOV)
  =& \tfrac{3}{2} \big[(\sigmaOV_1\!\cdot\aOV)(\sigmaOV_2\!\cdot\bOV) 
     + (\sigmaOV_1\!\cdot\bOV)(\sigmaOV_2\!\cdot\aOV)\big] \\
  &- \tfrac{1}{2}(\sigmaOV_1\!\cdot\!\sigmaOV_2) (\aOV\cdot\bOV +
    \bOV\cdot\aOV) \;.
}
Note, the tensor operator $\tensorRQO$ entering into the generator
$\gO_{\Omega}$ has the same structure as the standard tensor operator
$\tensorO = \tensorRRO$ appearing in the bare potential except for  the
replacement of one of the relative coordinate vectors by the orbital
momentum. Similar to the central correlators the functions
$\vartheta_T(r)$ describe the distance dependence of this angular shift for
isospin $T=0$ and $T=1$. Both,  $s_{ST}(r)$ and $\vartheta_T(r)$ have to be
in accord with the potential under consideration. 

The crucial difference between the Unitary Correlation Operator Method and
other schemes using similarity transformations to construct an effective
interaction, such as the Lee-Suzuki transformation \cite{SuLe80} or the Unitary
Model Operator Approach \cite{FuOk04}, is that our unitary correlation
operator is given in an explicit operator form. This enables us to
evaluate correlated wave functions or correlated operators analytically 
as will be shown in the following.

\subsection{Correlated Wave Functions}
\label{sec:corr_wavefunc}

We consider the effect of the correlation operators on the component of a 
two-nucleon state that describes the relative motion. The  center of mass
part is not affected by the unitary correlators because they depend only
on relative positions and momenta. For the uncorrelated relative wave
function we assume $LS$-coupled angular momentum eigenstates $\ket{\phi
(LS)J M\; T M_T}$. For the sake of simplicity, the projection quantum
numbers $M$ and $M_T$ are omitted in the following.

The central correlator $\cO_{r}=\exp(-\ii\, \gO_{r})$ 
\footnote{Small (capital) letters are used for correlation operators in two-body
($A$-body) space.}
affects only the radial part of the state
and leaves the angular momentum and spin components unchanged. In coordinate
representation it resembles a norm-conserving coordinate transformation 
\cite{FeNe98}
\eqmulti{
  \matrixe{r}{\cO_r}{\phi}
  &= \frac{\Rm(r)}{r}\sqrt{\DRm(r)}\; \braket{\Rm(r)}{\phi} \\
  \matrixe{r}{\ccO_r}{\phi}
  &= \frac{\Rp(r)}{r}\sqrt{\DRp(r)}\; \braket{\Rp(r)}{\phi} \;,
}
where $\Rp(r)$ and $\Rm(r)$ are mutually inverse, $\Rpm[\Rmp(r)]=r$. These
correlation functions are related to the function $s(r)$ in the generator  
\eqref{eq:central_generator} through the integral equation
\eq{
  \int_{r}^{\Rpm(r)} \frac{\dd{\xi}}{s(\xi)} = \pm 1 \;.
}
To a certain approximation the following intuitive relation holds 
$\Rpm(r) \approx r \pm s(r)$. For the sake of brevity we omit the spin
and isospin indices of the correlation functions here and in the following.

The action of the tensor correlator $\cO_{\Omega}$ on $LS$-coupled
two-body states can be evaluated directly \cite{NeFe03}. The matrix
elements of the tensor operator $\tensorRQO$ for those states have only
off-diagonal contributions 
\eqmulti{
  &\matrixe{\phi(J\pm1,1)JT}{\tensorRQO}{\phi(J\mp1,1)JT} \\ 
  &\quad= \pm 3\ii \sqrt{J(J+1)} \;.
}
Within a subspace of fixed $J$ one can easily obtain the matrix
exponential and thus the matrix elements of the full tensor correlator
$\cO_{\Omega}$. On this basis we can construct explicit relations for the
tensor correlated two-body states. 

States with $L=J$ are invariant under transformation with the tensor
correlation operator
\eq{ \label{eq:corr_tensor_states1}
  \cO_{\Omega} \ket{\phi (J S) J T}
  = \ket{\phi (J S) J T} \;.
}
Only states with $L=J\pm1$ are susceptible to tensor correlations and
transform like
\eqmulti{ \label{eq:corr_tensor_states2}
  \cO_{\Omega} \ket{\phi (J\pm 1,1) J T}
  &= \cos\theta_J(\rO)\, \ket{\phi (J\pm 1,1) J T} \\
  &\mp\, \sin\theta_J(\rO)\, \ket{\phi (J\mp 1,1) J T}
}
with $\rO$ being the radial distance operator and
\eq{
  \theta_J(\rO) 
  = 3 \sqrt{J(J+1)}\; \vartheta(\rO) \;.
}   
The tensor correlator admixes a state with $\Delta L=\pm 2$ and changes
the radial wave function of both components depending on the tensor
correlation function $\vartheta(r)$.

\begin{figure}
\begin{center}
\includegraphics[width=0.48\textwidth]{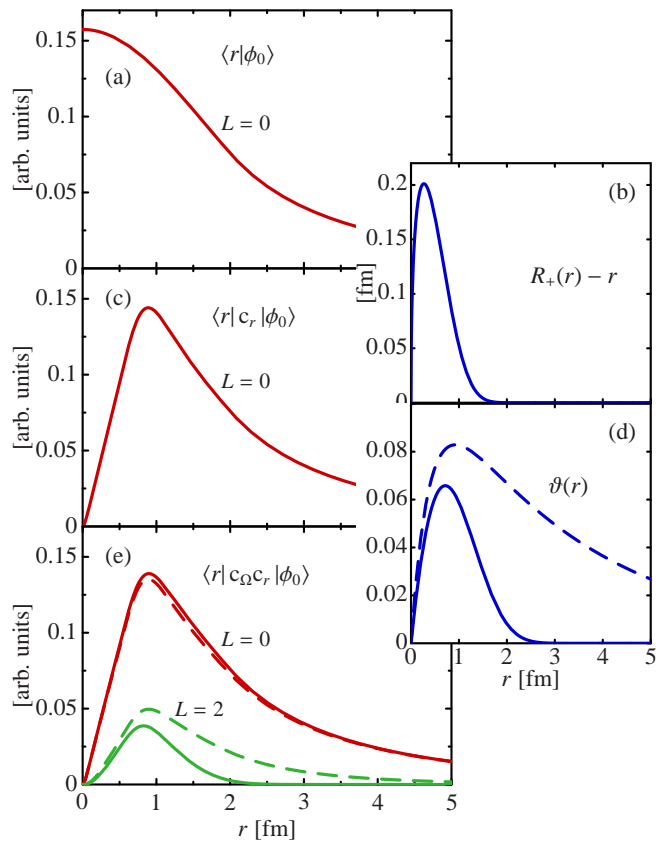}
\end{center}
\vskip-4ex
\caption{(Color online) Application of the central and tensor correlators to a 
deuteron-like two-body wave function. Panels (a), (c), and (e) depict the
uncorrelated, the central correlated, and the fully correlated radial wave
functions, resp. The panels (b) and (d) show the corresponding central and
tensor correlation functions (see text).}
\label{fig:deuteron_illu}
\end{figure}

To illustrate the impact of the central and tensor correlation operators
on a two-body state, we consider the example of the deuteron. Assume an
uncorrelated state $\ket{\phi_0(LS)JT} =\ket{\phi_0(01)10}$ which is a
pure $S$-wave state with the spin-isospin quantum numbers of the deuteron.
The radial wave function $\braket{r}{\phi_0}$ shall not contain
short-range correlations induced by the repulsive core. Fig.
\ref{fig:deuteron_illu}(a) shows the uncorrelated $L=0$ radial wave
function. Applying the central correlator $\cO_r$ with the correlation
function $\Rp(r)$ depicted in panel (b) leads to a wave function which has
the short-range central correlations, i.e. the hole at small interparticle
distances, built in as depicted in Fig. \ref{fig:deuteron_illu}(c). 
Subsequent application of the tensor correlator $\cO_{\Omega}$ with the
correlation function $\vartheta(r)$ depicted in panel (d) generates the
fully correlated wave function shown in Fig. \ref{fig:deuteron_illu}(e).
As a result of the tensor correlations, the wave function acquires a
$D$-wave admixture whose radial structure depends crucially on the tensor
correlation function. In order to represent the long-range D-wave
admixture, which is characteristic for the realistic deuteron wave function, 
a long-ranged tensor correlator is required as indicated by the
dashed curve in panels (d) and (e). In the following sections we will
argue that only the short-range and state-independent correlations should be
described by the correlation operator. The solid curves in panels (d) and
(e) correspond to an optimal short-range tensor correlator as will be
constructed later-on (see Sec. \ref{sec:optcorr}, optimal correlator for
$I_{\vartheta}=0.09\,\text{fm}^3$).

\subsection{Correlated Operators and Cluster Expansion}
\label{sec:clusterexp}

The explicit formulation of correlated wave functions for the many-body
problem becomes technically increasingly complicated and the equivalent
notion of correlated operators proves more convenient. 

The similarity transformation \eqref{eq:corr_operator} of an arbitrary
operator $\AO$ leads to a correlated operator which contains irreducible
contributions to all particle numbers. We can formulate a cluster
expansion of the correlated operator,
\eq{ \label{eq:clusterexp}
  \corr{\AO}
  = \CCO \AO \CO
  = \corr{\AO}^{[1]} + \corr{\AO}^{[2]} + \corr{\AO}^{[3]} + \cdots \;,
}
where $\corr{\AO}^{[n]}$ denotes the irreducible $n$-body part
\cite{FeNe98}. When starting with a $k$-body operator, all irreducible
contributions $\corr{\AO}^{[n]}$ with $n<k$ vanish. Hence, the unitary
transformation of a two-body operator --- the NN-interaction for
example --- yields a correlated operator containing a two-body contribution,
a three-body term, etc. 

The significance of the higher order terms depends on the range of the
central and tensor correlations \cite{RoNe04,NeFe03,FeNe98}. If
the range of the correlation functions is small compared to the mean
interparticle distance, then three-body and higher-order terms of the
cluster expansion are negligible. Discarding these higher-order
contributions leads to the two-body approximation 
\eq{
  \corr{\AO}^{C2}
  = \corr{\AO}^{[1]} + \corr{\AO}^{[2]} \;.
}
In principle, the higher-order contributions to the cluster expansion can be
evaluated systematically \cite{Roth00}. However, for many-body calculations
the inclusion of those terms is an extreme challenge.

Therefore, we restrict ourselves to the two-body approximation and choose
the correlation functions such that its applicability is ensured. As discussed
in detail in Sec. \ref{sec:ncsm} we can use exact solutions of the
many-body problem, e.g. in the no-core shell model framework, to estimate
the size of the omitted higher-order contributions.

\subsection{Correlated Hamiltonian -- Central Correlations}
\label{sec:corr_op_central}

In two-body approximation the unitary transformation of any relevant
operator with the central correlation operator can be evaluated
analytically. 

As a comprehensive example we consider a Hamiltonian consisting of kinetic
energy and a realistic NN-interaction. For convenience we assume  a
generic operator form of the interaction
\eq{ \label{eq:corr_op_potgeneric}
  \vO
  = \sum_p \frac{1}{2} [v_p(\rO) \OO_p + \OO_p v_p(\rO)]
}
with 
\eqmulti{ \label{eq:corr_op_potgenericop}
  \OO_p 
  = \{&1,\;\sigmasigmaO,\;\qO_r^2,\;\qO_r^2 \sigmasigmaO,\;
    \LOV^2,\;\LOV^2 \sigmasigmaO,\\
    &\spinorbitO,\;\tensorRRO,\;\tensorLLO\} \otimes \{1,\; \tautauO\} \;.
}
In order to accommodate momentum dependent terms, as they appear,
e.g., in the Nijmegen \cite{StKl94} or Bonn A/B potentials \cite{Mach89},
we have chosen an explicitly symmetrized form. Notice, that any quadratic
momentum dependence can  be expressed by the $\qO_r^2$ and $\orbitsqrO$
terms contained in \eqref{eq:corr_op_potgeneric}. For simplicity, charge
dependent terms are not explicitly discussed here. Nevertheless, they will
be included in Sec. \ref{sec:ncsm}.
 
For the formulation of the correlated Hamiltonian in two-body approximation,
it is sufficient to consider the Hamiltonian for a two-nucleon system given by 
\eq{
  \hO 
  = \TO + \vO
  = \tO_{cm} + \tO_{r} + \tO_{\Omega} + \vO \;,
}  
where we have decomposed the kinetic energy operator $\TO$ into a center of
mass contribution $\tO_{cm}$ and a relative contribution which in turn is
written as a sum of a radial and an angular part
\eq{
  \tO_{r}
  = \frac{1}{2\mu} \qO_r^2 \;,\quad
  \tO_{\Omega}
  = \frac{1}{2\mu} \frac{\orbitsqrO}{\rO^2} \;.
}
Applying the central correlator $\cO_r$ in two-body space leads to a 
correlated Hamiltonian consisting of the bare kinetic energy $\TO$ and
two-body contributions for the correlated radial and angular kinetic
energy, $\corr{\tO}_r^{[2]}$ and $\corr{\tO}_{\Omega}^{[2]}$, resp., as
well as the correlated two-body interaction $\corr{\vO}^{[2]}$
\eq{
  \ccO_r\, \hO\, \cO_r
  = \TO +  \corr{\tO}_r^{[2]} + \corr{\tO}_{\Omega}^{[2]} 
    + \corr{\vO}^{[2]} \;.
}

The explicit operator form of the correlated terms can be derived from a
few basic identities. The similarity transformation for the relative
distance operator $\rO$ results in the operator-valued function $\Rp(\rO)$
\eq{
  \ccO_r\, \rO\,\cO_r
  = \Rp(\rO) \;.
}
The unitarity $\ccO_r=\cO_r^{-1}$ implies that an arbitrary function of
$\rO$ transforms as
\eq{ \label{eq:corr_op_funcr}
  \ccO_r\, f(\rO)\,\cO_r
  = f(\ccO_r\, \rO\,\cO_r)
  = f(\Rp(\rO)) \;.
}
The interpretation of the unitary transformation in terms of a
norm-conserving coordinate transformation $r \mapsto \Rp(r)$ is evident.
For the radial momentum operator $\qO_r$ one finds the following
correlated form \cite{FeNe98}
\eq{ \label{eq:corr_op_reldist}
  \ccO_r\, \qO_r\,\cO_r
  = \frac{1}{\sqrt{\DRp(\rO)}}\;\qO_r\;\frac{1}{\sqrt{\DRp(\rO)}} \;.
}
With this, we obtain the following expression for the square of the radial 
momentum, which enters into the radial part of the kinetic energy
\eq{ \label{eq:corr_op_relmomsqr}
  \ccO_r\, \qO_r^2\,\cO_r
  = \frac{1}{2} \Big[\frac{1}{\DRp(\rO)^2} \qO_r^2 + \qO_r^2
    \frac{1}{\DRp(\rO)^2} \Big] + w(\rO)
}
with an additional local term depending only on the correlation function
$\Rp(r)$
\eq{
  w(r) 
  = \frac{7 \DDRp(r)^2}{4 \DRp(r)^4} - \frac{\DDDRp(r)}{2\DRp(r)^3} \;.
}   
All other basic operators, such as $\orbitsqrO$, $\spinorbitO$, $\tensorO$
etc. commute with the correlation operator $\cO_r$ and are therefore
invariant.

Based on these elementary relations we can explicitly construct the
two-body contributions to the correlated kinetic energy. For the radial
part we obtain using \eqref{eq:corr_op_relmomsqr}
\eqmulti{ \label{eq:corr_op_kinetic_rad}
  \corr{\tO}_r^{[2]}
  &= \ccO_r \tO_r \cO_r - \tO_r \\
  &= \frac{1}{2} \Big(\frac{1}{2\mu_r(\rO)} \qO_r^2 + \qO_r^2 \frac{1}{2\mu_r(\rO)}\Big)
    + \frac{1}{2\mu} w(\rO)
}
with a distance-dependent effective mass term
\eq{
  \frac{1}{2\mu_r(r)} 
  = \frac{1}{2\mu} \Big( \frac{1}{\DRp(r)^2} - 1 \Big) \;.
}
The two-body contribution to the correlated angular part of the kinetic
energy involves only the basic relation \eqref{eq:corr_op_reldist} and
reads
\eq{ \label{eq:corr_op_kinetic_ang}
  \corr{\tO}_{\Omega}^{[2]}
  = \ccO_r \tO_{\Omega} \cO_r - \tO_{\Omega}
  = \frac{1}{2\mu_{\Omega}(\rO)} \frac{\orbitsqrO}{\rO^2}
}
with a distance-dependent angular effective mass term	
\eq{
  \frac{1}{2\mu_{\Omega}(r)} 
  = \frac{1}{2\mu} \Big( \frac{r^2}{\Rp(r)^2} - 1 \Big) \;.
}		   

The momentum dependent terms of the NN-interaction
\eqref{eq:corr_op_potgeneric} transform in a similar manner like the
kinetic energy. Using \eqref{eq:corr_op_reldist} and
\eqref{eq:corr_op_relmomsqr} we obtain
\eqmulti{
  \ccO_r\; \frac{1}{2}\Big(& \qO_r^2 v(\rO) + v(\rO) \qO_r^2 \Big) \; \cO_r
    = \\
  =& \frac{1}{2}\Big( \frac{v(\Rp(\rO))}{\DRp(r)^2} \qO_r^2 
      + \qO_r^2 \frac{v(\Rp(\rO))}{\DRp(r)^2} \Big) \\
  &+ v(\Rp(\rO))\; w(\rO) - v'(\Rp(\rO)) \frac{\DDRp(r)}{\DRp(r)^2} \;.
}  
For all other terms of the NN-interaction \eqref{eq:corr_op_potgeneric}
the commutator relations $[\qO_r, \OO_p]=[\rO,\OO_p]=0$ are fulfilled and the
similarity transformation with the central correlator reduces to
\eq{  \label{eq:corr_op_localpot}
  \ccO_r\; v(\rO)\,\OO_p\; \cO_r
  = v(\Rp(\rO))\, \OO_p \;.
}
Many of the other relevant operators, e.g. the quadratic radius or
transition operators, can be transformed just as easily.

\subsection{Correlated Hamiltonian -- Tensor Correlations}
\label{sec:corr_op_tensor}

The transformation of the Hamiltonian with the tensor correlation operator
$\cO_{\Omega}$ is more involved. In general, it can be evaluated via the
Baker-Campbell-Hausdorff expansion 
\eqmulti{ \label{eq:corr_tensor_bch}
  \ccO_{\Omega}\, \AO\, \cO_{\Omega}
  &= \AO + \ii [\gO_{\Omega}, \AO] 
    + \frac{\ii^2}{2} [\gO_{\Omega},[\gO_{\Omega}, \AO]] 
    + \cdots \;.
}
Evaluation of the iterated commutators in some cases results in a
termination of the series expansion.  A trivial case is the distance
operator $\rO$ which commutes with the tensor generator $\gO_{\Omega}$ and
is thus invariant under the transformation 
\eq{
  \ccO_{\Omega}\, \rO\, \cO_{\Omega} 
  = \rO \;.
}
For the radial momentum operator $\qO_r$, the expansion
\eqref{eq:corr_tensor_bch} terminates after the first order commutators and we 
obtain the simple expression 
\eq{
  \ccO_{\Omega}\, \qO_r\, \cO_{\Omega} 
  = \qO_r - \vartheta'(\rO)\,\tensorRQO \;.
}
Likewise, we find for the tensor correlated quadratic radial momentum operator
\eqmulti{ \label{eq:corrT_op_momentumsqr}
  \ccO_{\Omega}\, \qO_r^2\, \cO_{\Omega}
  =& \qO_r^2 - [\vartheta'(\rO)\, \qO_r + \qO_r\, \vartheta'(\rO)]\, \tensorRQO \\
  &+ [\vartheta'(\rO)\,\tensorRQO]^2 \;,
}
where $\tensorRQO^2 = 9 [\SOV^2 + 3\spinorbitO + \spinorbitO^2$].
For all other operators of the interaction \eqref{eq:corr_op_potgeneric},
that involve angular degrees of freedom, the Baker-Campbell-Hausdorff
series does not terminate. Through the commutators additional tensor operators 
are generated. For example, the relevant first order commutators are  
\eqmulti{
  \comm{\gO_{\Omega}}{\tensorRRO}
  &= \ii \vartheta(\rO) [-24\, \PiO_1 - 18\, \spinorbitO + 3\, \tensorRRO] \\
  \comm{\gO_{\Omega}}{\spinorbitO}
  &= \ii \vartheta(\rO) [-\tensorbarQQO] \\
  \comm{\gO_{\Omega}}{\orbitsqrO}
  &= \ii \vartheta(\rO) [2\,\tensorbarQQO] \\
  \comm{\gO_{\Omega}}{\tensorLLO}
  &= \ii \vartheta(\rO) [7\,\tensorbarQQO] \;,
}
where
\eq{
  \tensorbarQQO
  = 2\rO^2 \tensorQQO + \tensorLLO - \tfrac{1}{2}\,\tensorRRO \;.
}
The next order generates higher powers of the orbital angular momentum
operator, e.g. an $\orbitsqrO\spinorbitO$ term, in addition. The resulting
accumulation of new operators enforces a truncation of the
Baker-Campbell-Hausdorff expansion at some finite order for the operator
representation \cite{RoNe04}. The basis representation introduced in Sec.
\ref{sec:corrme} does not require this approximation.

\subsection{Correlated Interaction -- $\VO_{\UCOM}$}
\label{sec:corr_VUCOM}

Subtracting the uncorrelated kinetic energy operator from the central and
tensor correlated Hamiltonian in two-body space defines the correlated
interaction $\vO_{\UCOM}$ in two-body approximation:
\eq{
  \vO_{\UCOM} 
  = \ccO_{r} \ccO_{\Omega} \hO \cO_{\Omega} \cO_{r} - \TO \;.
}

If we start from a realistic interaction which is given in an
operator representation, e.g. the AV18 potential, then the
correlated interaction also has a closed operator representation
\eq{ \label{eq:VUCOM}
  \vO_{\UCOM}
  = \sum_{p} \frac{1}{2} 
    \big[ \corr{v}_{p}(\rO)  \corr{\OO}_p + \corr{\OO}_p \corr{v}_{p}(\rO) 
    \big] \;,
} 
where
\eqmulti{ \label{eq:VUCOM_ops}
  \corr{\OO}_p 
  = \{&1,\;\sigmasigmaO,\;\qO_r^2,\;\qO_r^2 \sigmasigmaO,\;
    \LOV^2,\;\LOV^2 \sigmasigmaO,\\
    &\spinorbitO,\;\tensorRRO,\;\tensorLLO,\\
    &\tensorbarQQO,\;\qO_r\, \tensorRQO,\;\orbitsqrO\spinorbitO,\\
    &\orbitsqrO \tensorbarQQO,\dots \}  \otimes \{1,\; \tautauO\} \;.
}
The dots indicate that higher-order contributions of the
Baker-Campbell-Hausdorff expansion for the tensor transformation have been
omitted. The terms shown above result from a truncation of the expansion
\eqref{eq:corr_tensor_bch} after the third order in $\gO_{\Omega}$. For
most applications the inclusion of these terms is sufficient \cite{RoNe04}.

The existence of an operator representation of $\vO_{\UCOM}$ is essential
for many-body models which are not based on a simple oscillator or
plane-wave basis. One example is the Fermionic Molecular Dynamics model
\cite{FeSc00,Feld90} which uses a non-orthogonal Gaussian basis and
does not easily allow for a partial wave decomposition of the relative
two-body states. Nevertheless, it is possible to evaluate the two-body
matrix elements of $\vO_{\UCOM}$ analytically, which facilitates efficient
computations with this extremely versatile basis
\cite{RoNe04,NeFe04,NeFe05}.

As we have emphasized already, the operators of \emph{all observables}
have to be transformed \emph{consistently}. The unitary transformation of 
observables like quadratic radii, densities, momentum distributions, or
transition matrix elements is straightforward given the toolbox acquired
for the transformation of the Hamiltonian. The Unitary Correlation
Operator Method owes this simplicity to the explicit  state and
representation-independent form of the correlation operators.  In
contrast, in many other approaches for the construction of an effective
interaction, e.g. the Lee-Suzuki transformation
\cite{SuLe80,NaVa00b,CaNa02} or  the $V_{\text{low}k}$ renormalization
group method \cite{BoKu03}, the consistent derivation of effective
quantities other than the interaction is a complicated and rarely
addressed problem \cite{StBa05}. 

An important feature of $\vO_{\UCOM}$ results from the finite range
of the correlation functions $s_{ST}(r)$ and $\vartheta_T(r)$ entering
into the generators. Since the correlation functions vanish at large
distances --- i.e., the correlation operator acts as a unit operator at
large $r$ --- asymptotic properties of a two-body wave function are
preserved. This implies that $\vO_{\UCOM}$ is by construction
\emph{phase-shift equivalent} to the original NN-interaction. The unitary
transformation can, therefore, be viewed as a way to construct an infinite
manifold of realistic potentials, which all give identical phase-shifts. 

It is interesting to observe in which way the unitary transformation
changes the operator of the interaction while preserving the
phase-shifts.  The central correlator reduces the short-range repulsion in
the local part of the interaction and, at the same time, creates a
non-local repulsion  through the momentum-dependent terms. The tensor
correlator removes some strength from the local tensor interaction and
creates additional central contributions as well as new non-local tensor
terms.  Hence, the unitary transformation exploits the freedom to
redistribute strength between local and non-local parts of the potential
without changing the phase-shifts. The non-local tensor terms establish an
interesting connection to the CD Bonn potential, which among the realistic
potentials is the only one including non-local tensor contributions
\cite{MaSl01}.

\section{Correlated Two-body Matrix Elements}
\label{sec:corrme}

Having introduced the basic formalism of the Unitary Correlation Operator
Method, we can now derive two-body matrix elements of the correlated
interaction $\vO_{\UCOM}$. We consider relative  $LS$-coupled states of
the form $\ket{n (LS) J M \; T M_T}$, with a generic radial quantum number $n$,
relative orbital angular momentum $L$, spin $S$, total angular momentum
$J$, and isospin $T$. The matrix elements of $\vO_{\UCOM}$ thus read 
\eqmulti{ \label{eq:corr_me_VUCOM}
  &\matrixe{n(LS)J M T M_T}{\vO_{\UCOM}}{n'(L'S)J M T M_T} \\
  &= \matrixe{n(LS)J M T M_T}{\ccO_{r}\ccO_{\Omega}\,\hO\,\cO_{\Omega}\cO_{r} 
    - \TO}{n'(L'S)J M T M_T} \;.
}
The center of mass part of the two-body states is irrelevant for the
unitary transformation, since the correlation operator only acts on the
relative degrees of freedom of the two-body states. In the following
derivations the projection quantum numbers $M$ and $M_T$ are omitted for
simplicity. The formal framework discussed in the following is completely
independent of the particular choice of basis, only the angular momentum
structure is relevant. 

The Unitary Correlation Operator Method offers different ways to compute
these matrix elements. If we assume a NN-interaction of the form
\eqref{eq:corr_op_potgeneric}, then we can use the operator
representation \eqref{eq:VUCOM} of $\vO_{\UCOM}$ and evaluate the matrix
elements directly. This approach is computationally quite efficient. If
one expands the radial dependencies of the individual operator channels in
a sum of Gauss functions, all radial integrals can be calculated
analytically.  The matrix elements of the additional tensor operators
contained in $\vO_{\UCOM}$ can be given in closed form as well. However,
this direct approach relies on the truncation of the
Baker-Campbell-Hausdorff expansion \eqref{eq:corr_tensor_bch} employed to
evaluate the tensor correlation. 

In order to avoid this approximation for the tensor transformation we
apply the tensor correlators to the two-body states and make use of the
exact expressions \eqref{eq:corr_tensor_states1} and
\eqref{eq:corr_tensor_states2}.  The central correlators will be applied
to the operator as before, since a simple and exact expression for the
central correlated Hamiltonian exists (cf. Sec.
\ref{sec:corr_op_central}). We formally interchange the ordering of the
correlations operators using the identity
\eqmulti{
  \ccO_{r} \ccO_{\Omega}\, \hO\, \cO_{\Omega} \cO_{r} 
  &= (\ccO_{r} \ccO_{\Omega} \cO_{r})\, \ccO_{r}\, \hO\, \cO_{r}\, 
     (\ccO_{r} \cO_{\Omega} \cO_{r}) \\
  &=  \corr{\cO}^{\dag}_{\Omega} \ccO_{r} \, \hO\, \cO_{r} \corr{\cO}_{\Omega}
}
with the ``centrally correlated'' tensor correlation operator
\eq{
  \corr{\cO}_{\Omega}
  = \ccO_{r} \cO_{\Omega} \cO_{r}
  = \exp[-\ii\, \vartheta(\Rp(\rO))\, \tensorRQO] \;.
}
The central correlator commutes with $\tensorRQO$ and transforms therefore
only $\vartheta(\rO)$, see Eq. \eqref{eq:corr_op_funcr}. The tensor
correlator $\corr{\cO}_{\Omega}$ acts on $LS$-coupled two-body  states
with $L=J$ like the identity operator (cf. Sec. \ref{sec:corr_wavefunc})
\eq{
  \corr{\cO}_{\Omega} \ket{n(J S) J T}
  = \ket{n (J S) J T} \;.
}
For states with $L=J\pm1$ we have the simple relation
\eqmulti{
  \corr{\cO}_{\Omega} \ket{n(J\mp 1,1) J T} 
  &= \cos\corr{\theta}_J(\rO)\, \ket{n (J\mp 1,1) J T } \\
  &\pm \sin\corr{\theta}_J(\rO)\, \ket{n (J\pm 1,1) J T} \;,
}
where
\eq{
  \corr{\theta}_J(\rO) 
  = 3 \sqrt{J(J+1)}\, \vartheta(\Rp(\rO)) \;.
}   
Using these relations we can calculate the correlated two-body matrix 
elements exactly.

\subsection{Interactions in operator representation}

We first consider a bare potential given in the generic operator
representation \eqref{eq:corr_op_potgeneric} and derive the correlated
matrix elements for the local contributions of the form $v(\rO) \OO$ with
$[\rO,\OO] = [\qO_r,\OO] = 0$, which includes all operators of the set
\eqref{eq:corr_op_potgenericop} except for the $\qO_r^2$ terms. 

The matrix elements for $L=L'=J$ are not affected by the tensor
correlations, only the central correlators act according to
\eqref{eq:corr_op_localpot}.  In coordinate representation we obtain 
\eqmulti{ \label{eq:corr_me_local1}
  &\matrixe{n(JS)JT}{\ccO_{r} \ccO_{\Omega}\, v(\rO) \OO\, 
    \cO_{\Omega}\cO_{r}}{n'(JS)JT} = \\ 
  &= \int\!\dd{r}\,u_{n,J}^{\star}(r)\, u_{n',J}(r)\;
     \corr{v}(r)\; \matrixe{(JS)JT}{\OO}{(JS)JT} \;,
}
where $\corr{v}(r) = v(\Rp(r))$ is the transformed radial dependence of
the potential. The $u_{n,L}(r)= r \phi_{n,L}(r)$ are the radial relative wave 
functions of the oscillator basis or any other basis under consideration. 
For the diagonal matrix elements with $L=L'=J\mp 1$ we get
\eqmulti{ \label{eq:corr_me_local2}
  &\matrixe{n(J\mpS 1,1) J T}{\ccO_{r} \ccO_{\Omega}\, v(\rO) \OO\, 
    \cO_{\Omega}\cO_{r}}{n'(J\mpS 1,1) J T} = \\
  &= \int\!\dd{r}\,u_{n,J\mpS 1}^{\star}(r)\, u_{n',J\mpS 1}(r)\;
     \corr{v}(r) \\[-2pt]
  &\quad\times\big[\matrixe{(J\mpS 1,1) J T}{\OO}{(J\mpS 1,1) J T}\, 
    \cos^2 \corr{\theta}_J(r) \\ 
  &\quad+\;\, \matrixe{(J\pmS 1,1) J T}{\OO}{(J\pmS 1,1) J T}\, 
    \sin^2 \corr{\theta}_J(r) \\
  &\quad\pm\;\, \matrixe{(J\mpS 1,1) J T}{\OO}{(J\pmS 1,1) J T}\, 
    2 \cos\corr{\theta}_J(r) \sin\corr{\theta}_J(r) \big]
}
with $\corr{\theta}_J(r) = \theta_J(\Rp(r))$. Finally, the off-diagonal
matrix elements for $L=J\mp 1$ and $L'=J\pm 1$ read
\eqmulti{ \label{eq:corr_me_local3}
  &\matrixe{n(J\mpS 1,1) J T}{\ccO_{r} \ccO_{\Omega}\, v(\rO) \OO\, 
    \cO_{\Omega}\cO_{r}}{n'(J\pmS 1,1) J T} = \\  
  &= \int\!\dd{r}\, u_{n,J\mpS 1}^{\star}(r)\, u_{n',J\pmS 1}(r)\;
     \corr{v}(r) \\[-2pt]
  &\quad\times\big[ \matrixe{(J\mpS 1,1) JT}{\OO}{(J\pmS 1,1) JT}\,
    \cos^2 \corr{\theta}_J(\rO) \\
  &\quad-\;\, \matrixe{(J\pmS 1,1) JT}{\OO}{(J\mpS 1,1) JT}\,
    \sin^2 \corr{\theta}_J(\rO)\\
  &\quad\mp\;\, \matrixe{(J\mpS 1,1) JT}{\OO}{(J\mpS 1,1) JT}\,
    \cos\corr{\theta}_J(\rO) \sin\corr{\theta}_J(\rO)\\
  &\quad\pm\;\, \matrixe{(J\pmS 1,1) JT}{\OO}{(J\pmS 1,1) JT}\,  
    \sin\corr{\theta}_J(\rO) \cos\corr{\theta}_J(\rO) \big] \;.
}
Apart from the integration involving the radial wave functions, the
matrix elements of the operators $\OO$ in $LS$-coupled angular momentum
states are required. Only for the standard tensor operator
$\OO=\tensorRRO$ the off-diagonal terms on the right hand side of Eqs.
\eqref{eq:corr_me_local2} and \eqref{eq:corr_me_local3} contribute. For
all other operators in \eqref{eq:corr_op_potgenericop} the off-diagonal
matrix elements vanish, and the above equations simplify significantly. 

The effect of the tensor correlator is clearly visible in the structure
of the correlated matrix elements \eqref{eq:corr_me_local2} and
\eqref{eq:corr_me_local3}. It admixes components with $\Delta L = \pm2$ to
the states. Therefore, the correlated matrix element consists of a
linear combination of diagonal and off-diagonal matrix elements
$\matrixe{(LS)JT}{\OO}{(L'S)JT}$. In this way even simple operators, like
$\orbitsqrO$ or $\spinorbitO$ acquire non-vanishing off-diagonal
\emph{correlated} matrix elements \eqref{eq:corr_me_local3}. 

The momentum dependent terms of the potential
\eqref{eq:corr_op_potgeneric} allow for an exact evaluation of the
similarity transformation on the operator level. For the tensor correlated
form of the operator 
\eq{
  \vO_{qr}
  = \frac{1}{2}\big[v(\rO) \qO_r^2 + \qO_r^2 v(\rO)\big]
}
we obtain 
\eqmulti{
  \ccO_{\Omega} \vO_{qr} \cO_{\Omega} 
  &= \frac{1}{2}\big[v(\rO) \qO_r^2 + \qO_r^2 v(\rO)\big] 
    + v(\rO) [\vartheta'(\rO) \tensorRQO]^2 \\
  &- \big[v(\rO)\vartheta'(\rO)\qO_r  + \qO_r \vartheta'(\rO)v(\rO) \big] 
    \tensorRQO 
}
by using Eq. \eqref{eq:corrT_op_momentumsqr}. Subsequent inclusion of the
central correlations leads to the following expression for the diagonal
matrix elements with $L=L'=J$ in coordinate representation:
\eqmulti{ \label{eq:corr_me_momentum1}
  &\matrixe{n(JS) JT}{\ccO_{r} \ccO_{\Omega} \vO_{qr} \cO_{\Omega}\cO_{r}}
    {n'(JS) JT} \\
  &= \int\!\dd{r}\,\bigg\{ u_{n,J}^{\star}(r)\, u_{n',J}(r)\;
    \bigg[\corr{v}(\rO)\; w(\rO) - \corr{v'}(\rO)
    \frac{\DDRp(r)}{\DRp(r)^2}\bigg] \\
  &\quad-\frac{1}{2}\big[u_{n,J}^{\star}(r)\, u''_{n',J}(r) 
    +u_{n,J}^{\prime\prime\star}(r)\, u_{n',J}(r) \big] 
    \frac{\corr{v}(r)}{\DRp(r)^2} \bigg\} \;,
}
where $\corr{v'}(r) = v'(\Rp(r))$. As before, the tensor correlator does
not affect these matrix elements and only the central correlations lead to
a modification. For the diagonal matrix elements with $L=L'=J\mp1$ the
tensor terms contribute and we obtain
\eqmulti{ \label{eq:corr_me_momentum2}
  &\matrixe{n(J\mpS 1,1) JT}{\ccO_{r} \ccO_{\Omega} \vO_{qr}
     \cO_{\Omega}\cO_{r}}{n'(J\mpS 1,1) JT} \\
  &= \int\!\dd{r}\,\bigg\{ u_{n,J\mpS 1}^{\star}(r)\, u_{n',J\mpS 1}(r) 
    \bigg[\corr{v}(r)\; w(r) + \corr{v}(r)\;\corr{\theta'}_J(r)^2 \\[-2pt]  
  &\quad- \corr{v'}(r) \frac{\DDRp(r)}{\DRp(r)^2}\bigg] 
    -\frac{1}{2} \big[u_{n,J\mpS 1}^{\star}(r)\, u''_{n',J\mpS 1}(r) \\
  &\quad + u_{n,J\mpS 1}^{\prime\prime\star}(r)\, u_{n',J\mpS 1}(r) \big] 
    \frac{\corr{v}(r)}{\DRp(r)^2} \bigg\}
}
with $\corr{\theta'}_J(r) = \theta'_J(\Rp(r))$. Likewise, we find 
\eqmulti{ \label{eq:corr_me_momentum3}
  &\matrixe{n(J\mpS 1,1) JT}{\ccO_{r} \ccO_{\Omega} \vO_{qr}
    \cO_{\Omega}\cO_{r}}{n'(J\pmS 1,1) JT} \\
  &= \pm \int\!\dd{r}\,
    \big[u_{n,J\mpS 1}^{\star}(r)\, u'_{n',J\pmS 1}(r) 
     - u_{n,J\mpS 1}^{\prime\star}(r)\, u_{n',J\pmS 1}(r) \big] \\[-2pt]
  &\qquad\times  \frac{\corr{v}(r)\, \corr{\theta'}_J(r)}{\DRp(r)} 
}
for the off-diagonal matrix elements with $L=J\mp1$ and $L'=J\pm1$.

The matrix elements for the correlated radial and angular kinetic energy
can be constructed as special cases of the interaction matrix elements
discussed above. By setting $v(r) = 1/(2\mu_r(r))$ in Eqs.
\eqref{eq:corr_me_momentum1} to \eqref{eq:corr_me_momentum3} we obtain the
matrix elements for the effective mass part of the correlated radial
kinetic energy \eqref{eq:corr_op_kinetic_rad}. The matrix elements of the
additional local potential in \eqref{eq:corr_op_kinetic_rad} and the
angular kinetic energy \eqref{eq:corr_op_kinetic_ang} follow directly from
Eqs.  \eqref{eq:corr_me_local1} to \eqref{eq:corr_me_local3}.

\subsection{Interactions in partial-wave representation}

So far we have discussed interactions given in a closed operator
representation of the form \eqref{eq:corr_op_potgeneric}. However, many
modern interactions, e.g., the CD Bonn potential or recent chiral
potentials, are defined using a non-local partial-wave representation. 
This makes it difficult to employ them within many-body
models which do not allow for a partial-wave expansion of the two-body
states \cite{RoNe04}. Nevertheless, the calculation of central and tensor
correlated matrix elements of the form \eqref{eq:corr_me_VUCOM} is
straightforward for those interactions.

Consider a general non-local NN-potential in partial-wave representation. For
simplicity we assume the potential given in a generic coordinate 
space representation
\eqmulti{ \label{eq:pot_pwnonlocal}
  \vO 
  = &\int\!\dd{r}\,r^2 \int\!\dd{r'}\,r'^2 \sum_{L,L',S,J,T} \\ 
    &\ket{r (LS)JT}\; v_{LL'SJT}(r,r')\; \bra{r' (L'S)JT} \;,
}  
where $M$ and $M_T$ have been omitted for brevity. Interactions given in
momentum space can be easily transformed into this representation. 

For the construction of the correlated matrix elements we only need the
expressions for correlated two-body states used in the previous
section. For $L=L'=J$ the tensor correlations are not active and we obtain
\eqmulti{ \label{eq:corr_me_pwnonlocal1}
  &\matrixe{n(JS) J T}{\ccO_{r} \ccO_{\Omega}\, \vO\, \cO_{\Omega}\cO_{r}}
    {n' (JS) J T} \\ 
  &=\int\!\dd{r}\,r\,\RRp(r) \int\!\dd{r'}\,r'\,\RRp(r')\; 
   u_{n,J}^{\star}(r)\, u_{n',J}(r') \\
   &\quad\times \corr{v}_{J,J,S,J,T}(r,r') \;,
}
where $\corr{v}_{LL'SJT}(r,r') = v_{LL'SJT}(\Rp(r),\Rp(r'))$. Due to the
non-local character with respect to the relative coordinate, the metric
factors $\RRp(r)=\sqrt{\DRp(r)}\,\Rp(r)/r$ resulting from the
transformation of the radial wave function remain. For the diagonal matrix
elements $L=L'=J\mp 1$ of the non-local interaction
\eqref{eq:pot_pwnonlocal} we obtain 
\eqmulti{ \label{eq:corr_me_pwnonlocal2}
  &\matrixe{n(J\mpS 1,1) J T}
    {\ccO_{r} \ccO_{\Omega}\, \vO\, \cO_{\Omega}\cO_{r}}
    {n'(J\mpS 1,1) J T} = \\
  &= \int\!\dd{r}\,r\,\RRp(r) \int\!\dd{r'}\,r'\,\RRp(r')\; 
    u_{n,J\mpS 1}^{\star}(r)\, u_{n',J\mpS 1}(r') \\[-2pt]
  &\quad\times\big[ \corr{v}_{J\mpS 1,J\mpS 1,1,J,T}(r,r') 
    \cos\corr{\theta}_J(r) \cos\corr{\theta}_J(r') \\ 
  &\quad+\;\, \corr{v}_{J\pmS 1,J\pmS 1,1,J,T}(r,r') 
    \sin\corr{\theta}_J(r) \sin\corr{\theta}_J(r') \\
  &\quad\pm\;\, \corr{v}_{J\mpS 1,J\pmS 1,1,J,T}(r,r')  
    \cos\corr{\theta}_J(r) \sin\corr{\theta}_J(r') \\
  &\quad\pm\;\, \corr{v}_{J\pmS 1,J\mpS 1,1,J,T}(r,r')  
    \sin\corr{\theta}_J(r) \cos\corr{\theta}_J(r') \big]
}
with $\corr{\theta}_J(r) = \theta_J(\Rp(r))$. Finally, the off-diagonal
matrix elements with $L=J\mp1$ and $L'=J\pm1$ read
\eqmulti{ \label{eq:corr_me_pwnonlocal3}
  &\matrixe{n(J\mpS 1,1) J T}
    {\ccO_{r} \ccO_{\Omega}\, \vO\, \cO_{\Omega}\cO_{r}}
    {n'(J\pmS 1,1) J T} = \\
  &= \int\!\dd{r}\,r\,\RRp(r) \int\!\dd{r'}\,r'\,\RRp(r')\; 
    u_{n,J\mpS1}^{\star}(r)\, u_{n',J\pmS 1}(r') \\[-2pt]
  &\quad\times\big[ \corr{v}_{J\mpS 1,J\pmS 1,1,J,T}(r,r') 
    \cos\corr{\theta}_J(r) \cos\corr{\theta}_J(r') \\ 
  &\quad-\;\, \corr{v}_{J\pmS 1,J\mpS 1,1,J,T}(r,r') 
    \sin\corr{\theta}_J(r) \sin\corr{\theta}_J(r') \\
  &\quad\mp\;\, \corr{v}_{J\mpS 1,J\mpS 1,1,J,T}(r,r')  
    \cos\corr{\theta}_J(r) \sin\corr{\theta}_J(r') \\
  &\quad\pm\;\, \corr{v}_{J\pmS 1,J\pmS 1,1,J,T}(r,r')  
    \sin\corr{\theta}_J(r) \cos\corr{\theta}_J(r') \big] \;.
}

For local interactions $v_{LL'SJT}(r,r') =
v_{LL'SJT}^{\text{loc}}(r)\delta(r-r')/(r'r)$ the  metric factors 
$\RRp(r)$ can be eliminated and the above equations simplify
substantially. For technical reasons we use these expression also for the
calculations with the AV18 potential discussed in the following.

\section{Optimal Correlation Functions}
\label{sec:optcorr}

The Unitary Correlation Operator Method encapsulates the physics of 
short-range central and tensor correlations in the set of correlation 
functions $s(r)$ and $\vartheta(r)$. In this section, we discuss a scheme
to determine these correlation functions for a given NN-potential. One
important task is to isolate the short-range state-independent
correlations from residual long-range correlations that should not be
described by the unitary transformation but by the many-body state.

The most convenient procedure to fix the correlation functions is based
on an energy minimization in the two-body system \cite{NeFe03}. For each
combination of  spin $S$ and isospin $T$ we compute the correlated energy
expectation value using a two-body trial state with the lowest possible
orbital angular momentum $L$. The uncorrelated radial wave function
should not contain any of the short-range correlations, i.e., it should
resemble the short-range behavior of a non-interacting system. In the following we
will use a free zero-energy scattering solution $\phi_L(r)\propto r^L$.
One could just as well use harmonic oscillator wave functions, the
difference in the resulting correlation functions is marginal. 

The correlation functions are represented by parametrizations with
typically three variational parameters. The long-range part is generally 
well-described by a double-exponential decay with variable range. 
For the short-range behavior, several different parametrizations have been
compared. For the AV18 potential, the following two parametrizations for 
the central correlation functions have proven most appropriate:
\eqmulti{
  \Rp^{\text{I}}(r) 
  &= r + \alpha\, (r/\beta)^{\eta} \exp[-\exp(r/\beta)]  \;,\\
  \Rp^{\text{II}}(r)
  &= r + \alpha\, [1 - \exp(-r/\gamma)] \exp[-\exp(r/\beta)] \;.
}
Which of these parametrizations is best suited for a particular channel
will be decided on the basis of the minimal energy alone. Note that rather
than $s(r)$, we directly parametrize the function $\Rp(r)$, which enters
into the expressions for correlated operators and matrix elements. For the
tensor correlation functions the following parametrization is used 
\eq{
  \vartheta(r) 
  = \alpha\, [1 - \exp(-r/\gamma)] \exp[-\exp(r/\beta)] \;.
}

The $S=0$ channels are only affected by the central correlators. Their
parameters are determined from the energy minimization within the lowest
possible orbital angular momentum state, i.e. $L=1$ for $T=0$ and $L=0$ for
$T=1$, resp.,  
\eqmulti{
  E_{00} &= \matrixe{\phi_1 (10)1 0}{\ccO_r\, \hO\, \cO_r}{\phi_1 (10)1 0} \;,\\
  E_{01} &= \matrixe{\phi_0 (00)0 1}{\ccO_r\, \hO\, \cO_r}{\phi_0 (00)0 1} \;.\\
}  
For $S=0, T=1$ the minimization of $E_{01}$ by variation of the parameters
of the central correlation function is straightforward. The resulting
parameters are summarized in Table \ref{tab:corr_centralpara}. For $S=0,
T=0$ the potential is purely repulsive and, therefore, the energy
minimization, for a negligible gain in energy, leads to central
correlation functions of very long range. In order to avoid this pathology
we employ a constraint on the strength of the correlation function defined
through
\eq{ 
  I_{\Rp}
  = \int\dd{r}\, r^2\, (\Rp(r)-r) \;.
}
The value of this constraint on the central correlation function for the 
$S=0$, $T=0$ channel is fixed to $I_{\Rp}=0.1\text{fm}^{4}$ in accord with
the typical values in the other channels. 

\begin{table}[t]
\begin{ruledtabular}
\begin{tabular}{c c c c c c c}
$S$ & $T$ & Param. &  $\alpha$ [fm] & $\beta$ [fm] & $\gamma$ [fm] & $\eta$ \\
\hline
0 & 0 & II &  0.7971 &  1.2638  &  0.4621  &  ---     \\
0 & 1 & I  &  1.3793 &  0.8853  &  ---     &  0.3724 \\
1 & 0 & I  &  1.3265 &  0.8342  &  ---     &  0.4471 \\
1 & 1 & II &  0.5665 &  1.3888  &  0.1786  &  ---     \\
\end{tabular}
\end{ruledtabular}
\caption{Parameters of the central correlation functions $\Rp(r)$ for
the AV18 potential obtained from two-body energy minimization.}
\label{tab:corr_centralpara}
\end{table}
\begin{figure}
\begin{center}
\includegraphics[width=0.36\textwidth]{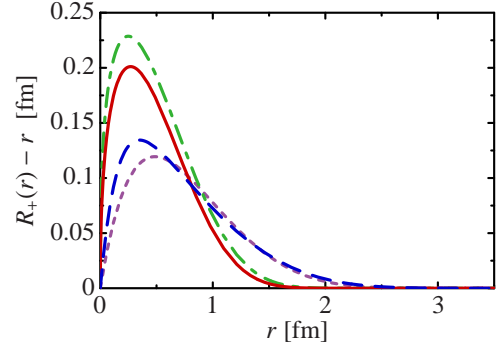}
\end{center}
\vskip-4ex
\caption{(Color online) Optimal central correlation functions $\Rp(r)-r$ for the AV18
potential according to the parameters given in Tab.
\ref{tab:corr_centralpara}. The curves correspond to the different
spin-isospin channels: $(S,T)=(0,1)$ \linemediumdashdot[FGGreen], 
  $(1,0)$ \linemediumsolid[FGRed], $(0,0)$ \linemediumdotted[FGViolet],
  $(1,1)$ \linemediumdashed[FGBlue]}
\label{fig:corr_central}
\end{figure}

For $S=1$ the tensor correlations are active as well and we
determine the parameters of the central and the tensor correlation
functions simultaneously. For $T=0$ the energy is defined by the matrix
element with $L=0$ states
\eq{
  E_{10} 
  = \matrixe{\phi_0 (01)1 0}{\ccO_r \ccO_{\Omega}\, \hO\, \cO_{\Omega}
      \cO_r}{\phi_0 (01)1 0} \;.
}
For $T=1$ the lowest possible orbital angular momentum is $L=1$. From
angular momentum coupling we obtain $0,1,$ and $2$ as possible values for
$J$. Therefore, we define the energy functional which is used in the
minimization procedure as the sum over all three possibilities with relative
weights given by $2J+1$
\eqmulti{
  E_{11}
  &= \tfrac{1}{9}\matrixe{\phi_1 (11)0 1}{\ccO_r\, \hO\, \cO_r}{\phi_1 (11)0 1} \\
    &+ \tfrac{3}{9}\matrixe{\phi_1 (11)1 1}{\ccO_r\, \hO\, \cO_r}{\phi_1 (11)1 1} \\
    &+ \tfrac{5}{9}\matrixe{\phi_1 (11)2 1}{\ccO_r\ccO_{\Omega}\, \hO\,
    \cO_{\Omega}\cO_r}{\phi_1 (11)2 1} \;.
}

\begin{table}
\begin{ruledtabular}
\begin{tabular}{c c c c c}
$T$ & $I_{\vartheta}$ [fm${}^3$] &  $\alpha$ [fm] & $\beta$ [fm] & $\gamma$ [fm] \\
\hline
0 & 0.03 &  491.32  &  0.9793  &  1000.0  \\
0 & 0.04 &  521.60  &  1.0367  &  1000.0  \\
0 & 0.05 &  539.86  &  1.0868  &  1000.0  \\
0 & 0.06 &  542.79  &  1.1360  &  1000.0  \\
0 & 0.07 &  543.21  &  1.1804  &  1000.0  \\
0 & 0.08 &  541.29  &  1.2215  &  1000.0  \\
0 & 0.09 &  536.67  &  1.2608  &  1000.0  \\
0 & 0.10 &  531.03  &  1.2978  &  1000.0  \\
0 & 0.11 &  524.46  &  1.3333  &  1000.0  \\
0 & 0.12 &  517.40  &  1.3672  &  1000.0  \\
\hline
1 & 0.01 &  -0.1036  &  1.5869  &  3.4426  \\
1 & 0.02 &  -0.0815  &  1.9057  &  2.4204  \\
1 & 0.03 &  -0.0569  &  2.1874  &  1.4761  \\
1 & 0.04 &  -0.0528  &  2.3876  &  1.2610  \\
1 & 0.05 &  -0.0463  &  2.6004  &  0.9983  \\
1 & 0.06 &  -0.0420  &  2.7984  &  0.8141  \\
1 & 0.07 &  -0.0389  &  2.9840  &  0.6643  \\
1 & 0.08 &  -0.0377  &  3.1414  &  0.6115  \\
1 & 0.09 &  -0.0364  &  3.2925  &  0.5473  \\
1 & 0.10 &  -0.0353  &  3.4349  &  0.4997  \\
\end{tabular}
\end{ruledtabular}
\caption{Parameters of the tensor correlation functions $\vartheta(r)$ for
the AV18 potential with different values $I_{\vartheta}$ for the
range constraint obtained from two-body energy minimization.}
\label{tab:corr_tensorpara}
\end{table}

As mentioned earlier, the long-range character of the tensor force leads
to long-range tensor correlations. However, long-range tensor correlation
functions are not desirable for several reasons: (\emph{i}) The optimal long-range
behavior would depend strongly on the nucleus under consideration. Hence,
our goal of extracting the state-independent, universal correlations
forbids long-range correlation functions. (\emph{ii}) The two-body approximation
would not be applicable for long-range correlators. (\emph{iii}) Effectively,
higher order contributions of the cluster expansion lead to a screening of
long-range tensor correlations between two nucleons through the presence
of other nucleons within the correlation range \cite{RoNe04}.  For these
reasons, we constrain the range of the tensor correlation functions in our
variational procedure.  We use the following integral constraint on the
``volume'' of the tensor correlation functions
\eq{
  I_{\vartheta}
  = \int \dd{r}\, r^2\; \vartheta(r) \;.
}
The constrained energy minimization for the $S=1, T=0$ and the $S=1, T=1$
channels with different values of the tensor correlation volume
$I_{\vartheta}$ leads to optimal parameters reported in Table
\ref{tab:corr_tensorpara}. The optimal parameters for the central
correlation functions change only marginally with the tensor constraint.
Therefore, we adopt a fixed set of parameters for the central correlators
given in Table \ref{tab:corr_centralpara}.

\begin{figure}[t]
\begin{center}
\includegraphics[width=0.36\textwidth]{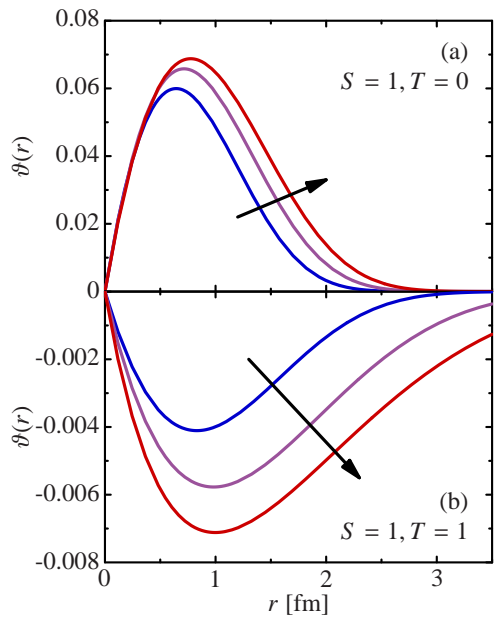}
\end{center}
\vskip-4ex
\caption{(Color online) Optimal tensor correlation functions $\vartheta(r)$ for different
values of the range constraint $I_{\vartheta}$. (a) Correlation functions
for $T=0$ with $I_{\vartheta}=0.06$, $0.09$, and
$0.12\,\text{fm}^3$. (b) Correlation functions
for $T=1$ with $I_{\vartheta}=0.01$, $0.03$, and
$0.06\,\text{fm}^3$. The arrows indicate the direction of increasing
$I_{\vartheta}$.}
\label{fig:corr_tensor}
\end{figure}

The optimal central correlation functions for the AV18 potential are
depicted in Fig. \ref{fig:corr_central}. In the even channels, the
correlation functions decrease rapidly and vanish beyond
$r\approx1.5\,\text{fm}$. The central correlators in the odd channels are
weaker and of slightly longer range due to the influence of the centrifugal
barrier. For the tensor correlation functions the constraints on the range
are important. Fig. \ref{fig:corr_tensor} shows the triplet-even (a) and
triplet-odd (b) tensor correlation functions $\vartheta(r)$ for different
$I_{\vartheta}$. Because the tensor interaction is significantly weaker
for $T=1$ than for $T=0$, the tensor correlator for this channel has a much
smaller amplitude. The relevant values for the constraint $I_{\vartheta}$
are therefore smaller for the triple-odd channel.

We stress that the range constraint for the tensor correlation functions
has an important physical and conceptual background. The Unitary
Correlation Operator Method is used to describe state-independent
short-range correlations only. Long-range correlations of any kind have to
be described by the model space employed in the solution of the many-body
problem. By constraining the range of the tensor correlators we set the
separation scale between short-range and long-range correlations. The
optimal value for tensor constraints cannot be fixed in the two-body system
alone, but requires input from few-nucleon systems. We will come back to this
point in Sec. \ref{sec:ncsm}.

\section{Properties of correlated momentum-space matrix elements}
\label{sec:momentumme}

\subsection{Effect of the Correlators}

\begin{figure*}
\begin{center}
\includegraphics[width=0.9\textwidth]{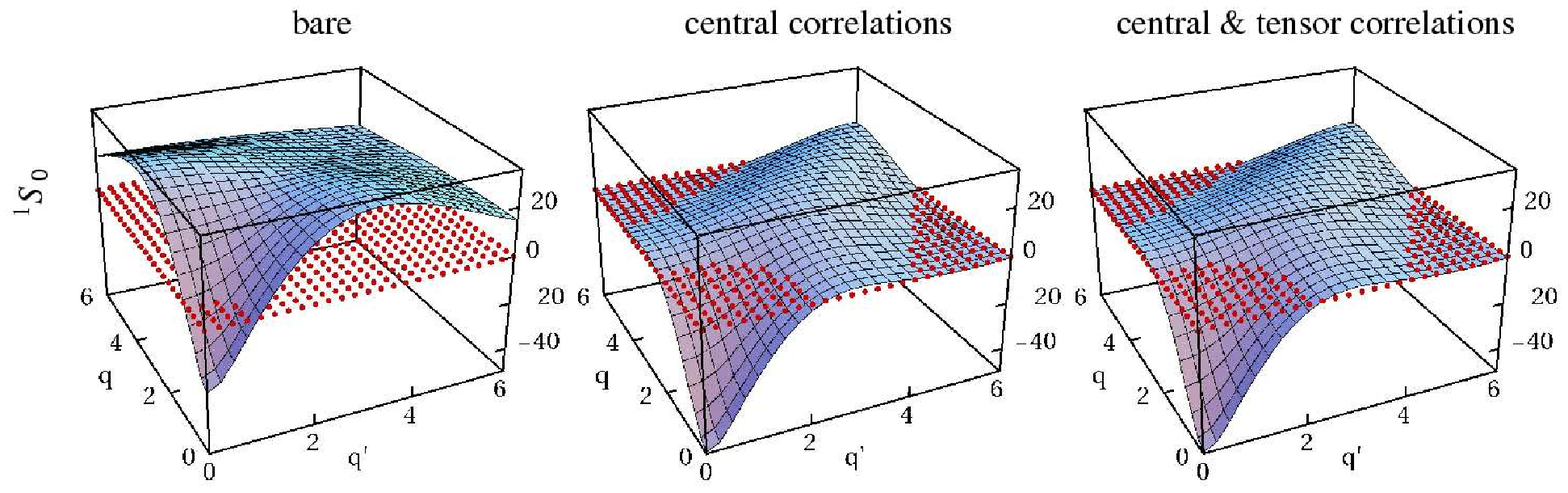} \\
\includegraphics[width=0.9\textwidth]{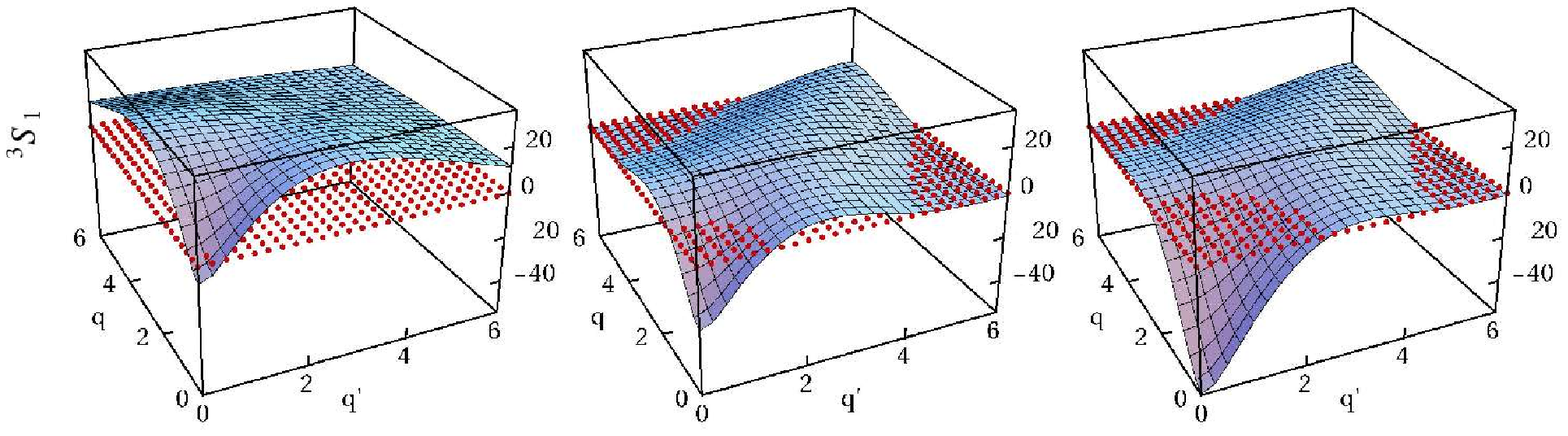} \\
\includegraphics[width=0.9\textwidth]{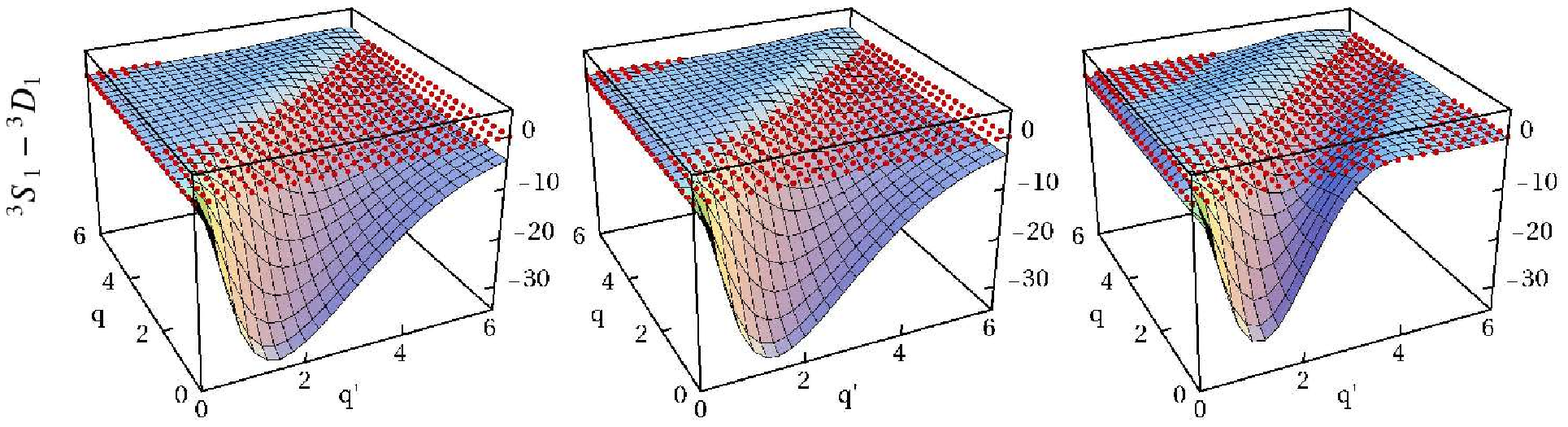} \\
\end{center}
\vspace*{-4ex}
\caption{(Color online) Relative momentum-space matrix elements of the bare AV18
potential (left-hand column), the central correlated AV18 potential
(center column), and the fully correlated AV18 potential. The
different rows correspond to different partial waves:  ${}^{1}S_{0}$
(top row), ${}^{3}S_{1}$ (middle row), and  ${}^{3}S_{1}-{}^{3}D_{1}$
(bottom row). The optimal tensor correlator for 
$I_{\vartheta}=0.09\,\text{fm}^3$ is used. The red dots mark the
plane of vanishing matrix elements. The momenta are given in units of
$[\text{fm}^{-1}]$ and the matrix elements in $[\text{MeV}]$.}
\label{fig:momentumme3d}
\end{figure*}

In order to illustrate the effect of the unitary transformation in more
detail, we discuss relative momentum space matrix elements of the form
$\matrixe{q(LS)JT}{\vO_{\UCOM}}{q'(L'S)JT}$, where $q$ is the relative
two-body momentum. The calculation of correlated momentum-space matrix
elements is performed using the relations derived in Sec. \ref{sec:corrme}
with radial wave functions given by the spherical Bessel functions. 

First, we consider the full set of matrix elements in the $(q,q')$-plane
for the lowest partial waves and compare the bare AV18 potential with the
correlated interaction. The plots in Fig. \ref{fig:momentumme3d} depict
the matrix elements for the ${}^{1}S_{0}$ and the ${}^{3}S_{1}$ partial
waves as well as for the mixed ${}^{3}S_{1}-{}^{3}D_{1}$ channel (from top
to bottom). The left-hand column corresponds to matrix elements of the
bare AV18 potential, the center column to correlated matrix elements using
the central correlator only, and the right-hand column to the fully
correlated matrix elements of $\vO_{\UCOM}$ including central and tensor
correlators.   

The gross effect of the unitary transformation on the dominant $S$-wave
matrix elements depicted in the upper two rows of Fig.
\ref{fig:momentumme3d} is similar. In both cases the matrix elements of
the bare interaction are predominantly repulsive except for a very small
region at small momenta. The inclusion of the central correlator,
accounting for correlations induced by the repulsive core of the
interaction, causes a substantial change in the correlated matrix
elements. In a region of low momenta $q, q' \lesssim 2\,\text{fm}^{-1}$
the matrix elements become strongly attractive. For larger momenta the
magnitude of the matrix elements is reduced, outside a band along the
diagonal the momentum space matrix elements practically vanish. Only
within this band a moderate repulsion remains. The inclusion of the tensor
correlator does not change the matrix elements in the spin-singlet
channel. In the spin-triplet channel the addition of tensor correlations
enhances the effect of the central correlations. The attractive matrix
elements at low momenta are enhanced while the off-diagonal matrix
elements are further suppressed. 

These matrix elements demonstrate the two major effects of the unitary 
correlators: (\emph{i}) For the important $L=0$ partial waves, the low-momentum
matrix elements become strongly attractive as a result of the proper
treatment of the correlations induced by the repulsive core and the tensor
part. (\emph{ii}) The off-diagonal matrix elements outside a band along the
diagonal are strongly suppressed. Hence the unitary transformation acts
like a \emph{pre-diagonalization}.

The importance of tensor correlations is accentuated in the
${}^{3}S_{1}-{}^{3}D_{1}$ channel depicted in the bottom row of Fig. 
\ref{fig:momentumme3d}. For the bare interaction only the tensor part
contributes to this channel and the matrix elements reveal strong
off-diagonal contributions. In many-body calculations, e.g. in a
shell-model framework, these off-diagonal matrix elements are responsible
for the admixture of high-lying basis states to the  ground state
contributing strongly to the binding energy \cite{VaSa73}. The effect of
the central correlator on these matrix elements is marginal. The tensor
correlator, however, causes a significant reduction of the off-diagonal
matrix elements. Outside of a band along the diagonal the matrix elements
vanish, i.e. the admixture of higher momenta or oscillator shells is
suppressed significantly.

The off-diagonal contributions from the tensor interaction are not fully
suppressed by the tensor correlators --- only the high-momentum components
are eliminated. This corresponds to the short-range part of the tensor
correlations in coordinate space, which we constrained the tensor
correlation operator to.

\begin{figure}
\begin{center}
\includegraphics[width=0.32\textwidth]{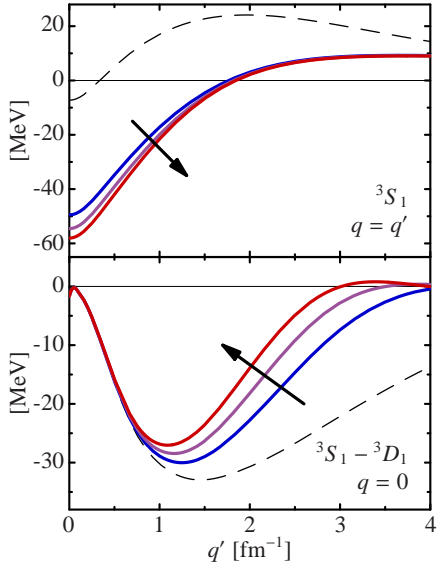}
\end{center}
\caption{(Color online) Momentum-space matrix elements of the correlated 
AV18 potential for different ranges of the tensor correlator
(solid lines): $I_{\vartheta}=0.06\,\text{fm}^{3}$, $0.09\,\text{fm}^{3}$,
$0.12\,\text{fm}^{3}$. The arrow indicates the direction of increasing
correlator range. The dashed line represents the matrix elements of the
bare potential. The upper panel shows diagonal matrix elements in the
${}^{3}S_1$ channel as function of $q=q'$. The lower panel depicts the
off-diagonal ${}^{3}S_1-{}^{3}D_1$ matrix elements for fixed $q=0$ as
function of $q'$. }
\label{fig:momentumme_range}
\end{figure}

The dependence of the momentum space matrix elements on the range of the
tensor correlation functions is illustrated in Fig.
\ref{fig:momentumme_range}. Note that only the tensor correlation functions
are changed, the central correlators stay the same. Therefore, only the
matrix elements in the spin-triplet channels change, and of those the $S=1,
T=0$ channels are affected most. The upper panel depicts the diagonal
$q=q'$ matrix elements for the ${}^{3}S_1$ channel. With increasing
correlator range $I_{\vartheta}$, as indicated by the arrow, the attraction
at low momenta is enhanced. This can be easily understood in the picture
of correlated states: Longer-ranged tensor correlators generate a
longer-range $D$-wave admixture such that the tensor attraction of the
bare potential can be exploited to a larger degree. In the picture of a
correlated Hamiltonian, the increased low-momentum attraction results from
a transformation of longer-ranged components of the tensor interaction
into operator channels which are accessible to uncorrelated $S$-wave
states (cf. Sec. \ref{sec:corr_VUCOM}). 

The off-diagonal matrix elements in the ${}^{3}S_1-{}^{3}D_1$ channel show
a complementary behavior. The lower panel in Fig.
\ref{fig:momentumme_range} depicts the off-diagonal matrix elements as 
functions of $q'$ for fixed $q=0$. As mentioned earlier, the matrix
elements far off the diagonal are strongly suppressed by the unitary
transformation---they are associated with short-range tensor correlations.
With increasing range of the tensor correlator, off-diagonal matrix
elements at successively lower momenta are suppressed as well. Hence the
band of non-vanishing matrix elements along the diagonal is narrowed with
increasing correlator range.

\subsection{Comparison with $V_{\text{low}k}$}

\begin{figure*}
\begin{center}
\includegraphics[width=0.9\textwidth]{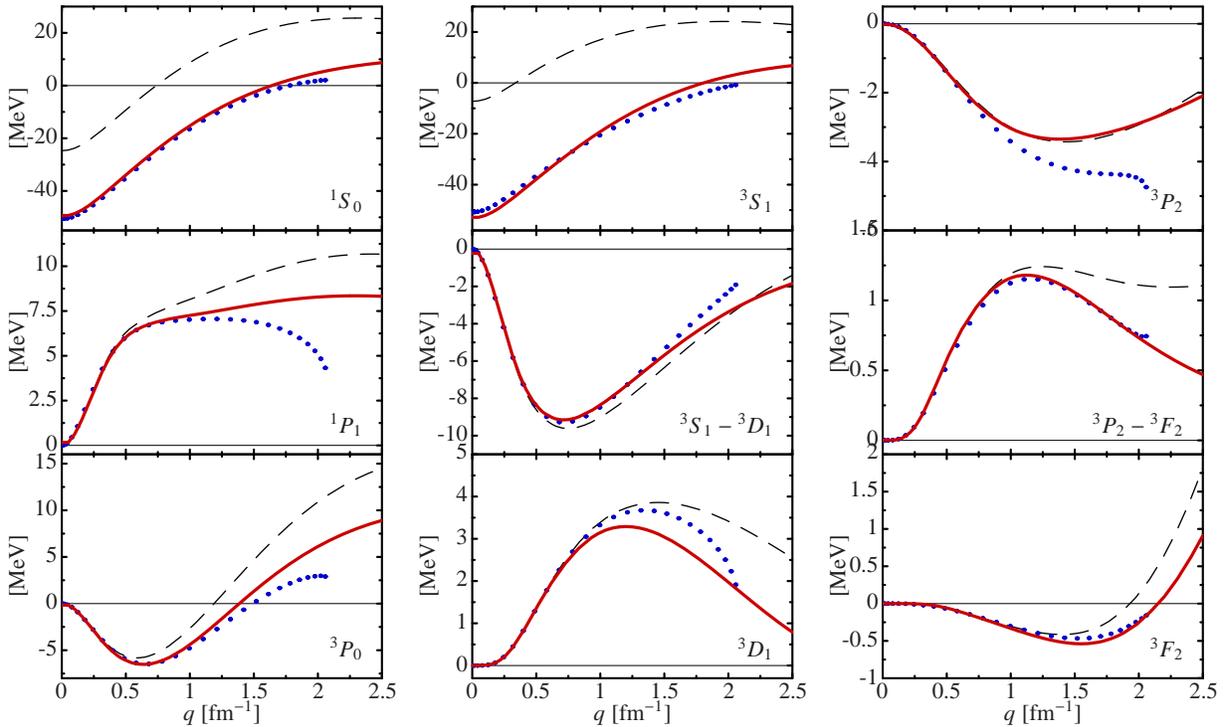}
\end{center}
\caption{(Color online) Comparison of the diagonal momentum space matrix elements 
of the correlated interaction $\vO_{\UCOM}$
(red lines) with the $V_{\text{low}k}$ matrix elements (blue dots) for the
AV18 potential. The dashed line corresponds to the matrix elements
of the bare potential. The optimal tensor correlators for
$I_{\vartheta}=0.08\,\text{fm}^3$ are used and the $V_{\text{low}k}$
momentum cutoff is $\Lambda=2.1\,\text{fm}^{-1}$.
}
\label{fig:momentumme_vlowk}
\end{figure*}

On the level of momentum-space matrix elements we can directly compare the
correlated interaction $\vO_{\UCOM}$ with the $V_{\text{low}k}$ matrix
elements resulting from a renormalization group decimation of the bare
interaction \cite{BoKu03,BoKu03b}. Both approaches aim at the construction
of a phase-shift equivalent low-momentum interaction, though their formal
background is completely different. The $V_{\text{low}k}$ approach relies
on a decoupling of a low-momentum $P$-space, constrained by a momentum
cutoff $\Lambda$, and a complementary high-momentum $Q$-space via a
similarity transformation. After a second transformation in order to
restore hermiticity, the momentum-space matrix elements within the
$P$-space are obtained. Matrix elements between $P$ and $Q$-space vanish
by virtue of the decoupling condition, $Q$-space matrix elements are
discarded (hence violating unitarity) such that nonvanishing matrix
elements exist only for momenta below the cutoff. In contrast to
$\vO_{\UCOM}$ the $V_{\text{low}k}$ approach is entirely formulated at the
level of momentum-space matrix elements for the different partial waves. 
This entails that a general operator representation of the effective
interaction is not directly accessible.

Despite their formal differences the matrix elements of both methods show
a remarkable agreement in the dominant partial waves. Fig.
\ref{fig:momentumme_vlowk} compares the  $V_{\text{low}k}$ matrix elements
for a cutoff momentum $\Lambda=2.1\,\text{fm}^{-1}$ with the
momentum-space matrix elements of $\vO_{\UCOM}$ obtained with the optimal
correlators for $I_{\vartheta}=0.08\,\text{fm}^3$. Up to momenta $q
\approx  1.5\,\text{fm}^{-1}$ the matrix elements agree very well in most
partial waves. For the dominant $S$-wave channels the agreement extends
right up to the cutoff momentum for $V_{\text{low}k}$. Above the cutoff
momentum the $V_{\text{low}k}$ matrix elements are zero by construction.
In contrast, the matrix elements of $\vO_{\UCOM}$ continuously extend to
larger momenta. This reflects the different conceptual ideas: Whereas
$V_{\text{low}k}$ attempts a decimation of the interaction to low-momentum
contributions below the cutoff scale, $\vO_{\UCOM}$ uses a
prediagonalization of the matrix-elements by a unitary transformation.

\section{Few-Body Shell-Model Calculations}
\label{sec:ncsm}

The correlated interaction and the correlated matrix elements can be used
as input for all kinds of many-body calculations. We have already
discussed nuclear structure studies in the framework of Fermionic
Molecular Dynamics \cite{RoNe04,NeFe03}, which rely on the operator form
of the correlated interaction. The correlated matrix elements serve as an
input for shell-model or Hartree-Fock calculations. In a forthcoming
publication we will present nuclear structure calculations based on
correlated realistic interactions in the framework of the Hartree-Fock and 
the Random Phase Approximation covering all mass regimes.

In this section we discuss the application of the correlated matrix
elements in no-core shell model calculations for \elem{H}{3} and
\elem{He}{4}. These few-body systems provide important information on the
correlated interaction beyond the two-body level.  We use the no-core
shell model code developed by Petr Navr\'atil et al.
\cite{NaKa00}. It is formulated in a translationally invariant harmonic
oscillator basis using Jacobi coordinates. Input for the shell-model
diagonalization are the relative two-body matrix elements of the
correlated AV18 potential, including charge dependent terms and  Coulomb
interaction. We stress that the Lee-Suzuki transformation usually 
employed in the no-core shell-model \cite{NaVa00b,CaNa02,NaKa00} is not
used here. We only perform a plain shell-model diagonalization. The task
of transforming the bare interaction into an effective interaction
suitable for shell-model calculations in small model-spaces is performed
by the unitary correlation operators. 

\begin{figure}
\begin{center}
\includegraphics[width=0.35\textwidth]{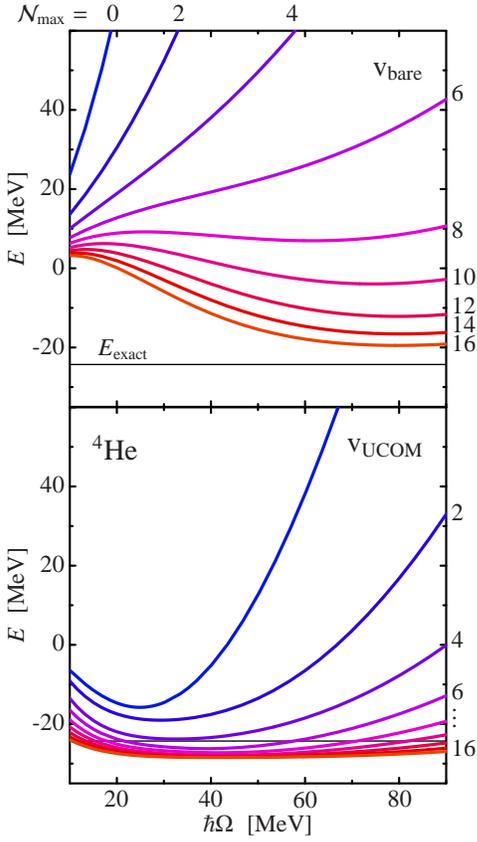}
\end{center}
\caption{(Color online) Ground state energy of $\elem{He}{4}$ as function of the
oscillator parameter $\hbar\Omega$ for different model model-space sizes
$\NC_{\max}=0,2,\dots,16$ as indicated by the
labels. The upper panel shows results for the bare AV18 potential,
the lower panel corresponds to the correlated potential $\vO_{\UCOM}$ for
$I_{\vartheta}=0.09\,\text{fm}^{3}$. The horizontal lines
represent the exact binding energy for the bare potential taken from 
\cite{NoKa00}.}
\label{fig:ncsm_covergence_illu}
\end{figure}

The effect of the unitary correlation operators in a no-core shell model
calculation for the ground state of \elem{He}{4} is illustrated in Fig.
\ref{fig:ncsm_covergence_illu}. For a given size of the model-space,
characterized by the maximum relative oscillator quantum number
$\NC_{\max} = 2N_{\max}+L_{\max}$, the ground state energy is plotted as a
function of the oscillator parameter $\hbar\Omega$. The upper
panel depicts the shell-model result for the bare AV18 potential
without any explicit treatment of correlations. All correlations induced by
the interaction have to be described by the degrees of freedom of the model
space alone. As expected, huge model spaces are required in order to
adequately describe short-range correlations. Within the computational
limits of $\NC_{\max} \leq 16$ for \elem{He}{4}, one is not able to achieve
convergence with the bare interaction. The exact \elem{He}{4} ground state
energy for the AV18 potential \cite{NoKa00} (marked by the horizontal line) 
is still somewhat lower than the result from the shell-model diagonalization for 
$\NC_{\max}=16$.

The convergence behavior changes dramatically once we use the correlated
matrix elements instead of the bare ones. The lower panel in Fig. 
\ref{fig:ncsm_covergence_illu} depicts the no-core shell model results for
\elem{He}{4} obtained with the correlated AV18 potential using the tensor 
correlator for $I_{\vartheta}=0.09\,\text{fm}^3$ in the dominant $S=1,
T=0$ channel. The $S=1,T=1$ channel is irrelevant for the nuclei
considered in this section and the corresponding tensor correlation
function is set to zero. The comparison with the calculation for the bare AV18
potential reveals three major effects of the unitary transformation: 

(\emph{i}) The ground state energy for very small  model spaces, e.g.
$\NC_{\max}=0$, for which the space consists of a single  Slater
determinant, is lowered dramatically. Evidently, the inclusion of the
dominant short-range central and tensor correlations through the unitary 
transformation is sufficient to reproduce the bulk of the binding energy. 

(\emph{ii}) With increasing model-space size the energy is lowered by a
moderate amount. The convergence is drastically improved. Fully
converged results, featuring a flat energy curve over a significant range
of oscillator parameters $\hbar\Omega$ can be obtained in spaces of moderate
size. The energy gain compared to the results in small model-spaces can be
attributed to residual \emph{long-range} correlations not described by the
unitary correlator. In contrast to \emph{short-range} correlations, these 
\emph{long-range} correlations can be described quite easily in
model-spaces of manageable size, hence the fast convergence. 

(\emph{iii}) The converged energy is generally below the exact ground
state energy for the potential under consideration. This violation of the
variational bound is solely due to the omission of the three- and
four-body terms in the cluster expansion \eqref{eq:clusterexp} of the
correlated Hamiltonian. As a direct consequence of the unitarity of the
transformation, the exact energy eigenvalues of the correlated Hamiltonian
including all terms of the cluster expansion are identical to the exact
energy eigenvalues of the bare Hamiltonian. Hence, the difference between
the exact result using the bare interaction and the correlated interaction
in two-body approximation equals the contribution of higher cluster orders
and thus provides a quantitative measure for the quality of the two-body
approximation.

\begin{figure}
\begin{center}
\includegraphics[width=0.35\textwidth]{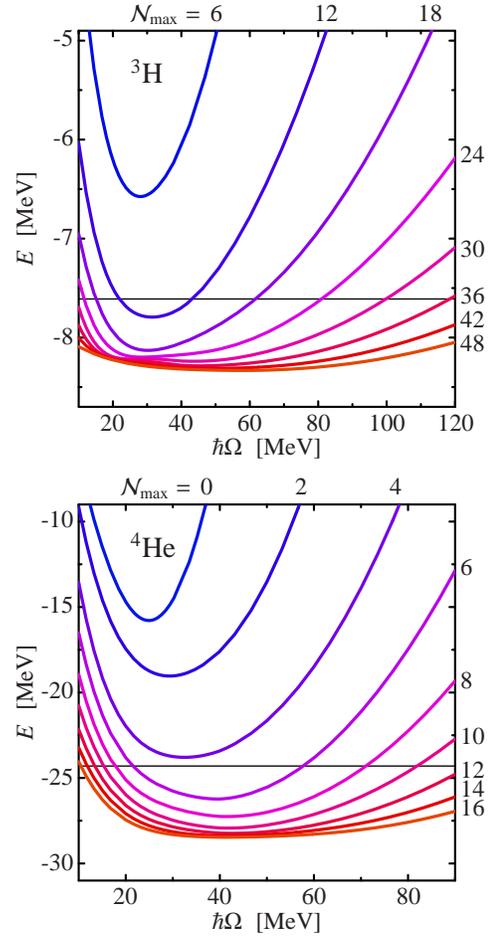}
\end{center}
\caption{(Color online) Ground state energy of $\elem{H}{3}$ and $\elem{He}{4}$ as function of the
oscillator parameter $\hbar\Omega$ for the correlated AV18 potential 
($I_{\vartheta}=0.09\,\text{fm}^{3}$) obtained in a no-core shell-model 
diagonalization. The different curves correspond to different model-space 
sizes $\NC_{\max}$ as indicated by the labels. The horizontal lines
represent the exact binding energies for the bare potential taken from 
\cite{NoKa00}.}
\label{fig:ncsm_covergence_detail}
\end{figure}

A more detailed view of the convergence behavior for the correlated AV18
potential is presented in Fig. \ref{fig:ncsm_covergence_detail} for
\elem{H}{3} and \elem{He}{4}. For both systems the three aforementioned
effects can be observed. The convergence for \elem{H}{3} is somewhat
slower because of the long-range structure of the wave function which
cannot easily be described within the oscillator basis. 

The no-core shell model results directly reflect the basic aims of the
Unitary Correlation Operator Method. The short-range correlations are
described explicitly by a state-independent unitary transformation such
that the bulk of the binding energy can readily be obtained in very small
model spaces. Residual system-dependent long-range correlations have to be
described by the model space, which is easily possible in the framework of
the no-core shell model. 

\begin{figure}
\begin{center}
\includegraphics[width=0.38\textwidth]{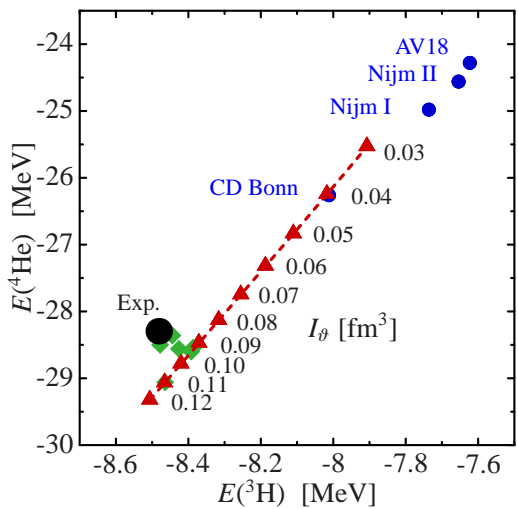}
\end{center}
\caption{(Color online) Binding energies of \elem{He}{4} versus \elem{H}{3}. The blue
discs show the results of exact Faddeev calculations obtained by A. Nogga
et al. \cite{NoKa00} using different modern NN-potentials. The green
diamonds show results obtained by including simple  three-body forces in
addition to the realistic two-nucleon potentials \cite{NoKa00}.  The red
triangles are the converged  no-core shell model results for the
correlated AV18 potential for different values
$I_{\vartheta}=0.03,\dots,0.12\,\text{fm}^3$ of the range-constraint for
the tensor correlator.}
\label{fig:ncsm_tjonline}
\end{figure}

Considering the omission of the higher-order terms in the cluster
expansions, the no-core shell model results allow for a study of the
range-dependence of these contributions. The difference between converged
result and exact calculations with the bare potential, i.e., the size of 
the omitted three- and four-body terms, increases with increasing range 
of the tensor correlators. This dependence is summarized in Fig. 
\ref{fig:ncsm_tjonline}, where we plot the ground state energies of
\elem{He}{4} and \elem{H}{3} in the $E(\elem{H}{3})$-$E(\elem{He}{4})$
plane. The results for different tensor correlators with range constraints
$I_{\vartheta}=0.03,\dots,0.12\,\text{fm}^3$ (cf. Tab.
\ref{tab:corr_tensorpara}) are represented by the triangles. All those 
points fall onto a straight line. Moreover, this line coincides with the
so-called Tjon-line \cite{NoKa00}, which characterizes a correlation
between the \elem{He}{4} and \elem{H}{3} binding energies found for
different realistic two-body potentials that reproduce the same phase
shifts (blue discs in Fig. \ref{fig:ncsm_tjonline}). It is not surprising
that the correlated interactions follow the same trend, because  the
correlated interaction $\vO_{\UCOM}$ generates the same phase-shifts as
the original potential. For each set of correlators, the resulting
correlated interaction $\vO_{\UCOM}$ provides a  new phase-shift
equivalent realistic potential. The value of the range constraint
$I_{\vartheta}$ can be used to map out the Tjon line.  A similar behavior
was observed for the $V_{\text{low}k}$ interaction as a function of the
cutoff parameters \cite{NoBo04}. 

In our calculation only the range of the tensor correlators in the
spin-triplet channels is varied and of those only the $S=1, T=0$ channel
is relevant for \elem{He}{4} and \elem{H}{3}. Hence, the variation along
the Tjon-line is related exclusively to the tensor correlations. As discussed
in Sec. \ref{sec:corr_op_tensor}, the unitary transformation of the
Hamiltonian with the tensor correlators produces additional
momentum-dependent tensor terms in $\vO_{\UCOM}$ as well as different
central contributions. The non-local tensor contributions seem to play an 
important role regarding the Tjon-line: Increasing the strength of the
non-local tensor contribution by increasing the strength of the tensor
correlators shifts the binding energies towards the experimental values and
away from the realistic potentials with purely local tensor contributions.
This is in accord with the results for the CD Bonn potential, which is the
only one among the high-precision NN-potentials including non-local
tensor contributions \cite{MaSl01}.

By choosing an appropriate value for the range constraint, one obtains a
phase-shift equivalent two-body potential, which produces binding energies
for \elem{He}{4} and \elem{H}{3} close to the experimental point
(cf. Fig. \ref{fig:ncsm_tjonline}). This is the result of a subtle
cancellation between different three-body contributions. For exact
calculations using the bare potentials one immediately finds that the
two-body potential alone (blue discs in Fig. \ref{fig:ncsm_tjonline}) does
not generate enough binding. It has to be supplemented by a three-body
force which produces a net attraction. The green diamonds in Fig. 
\ref{fig:ncsm_tjonline} show results obtained by A. Nogga et al.
\cite{NoKa00} using various simple parametrizations of phenomenological
three-nucleon forces supplementing the different realistic two-body
potentials. For the calculation with the correlated interactions we have
not included any of the three-body contributions, i.e., neither the
genuine three-body force nor the three-body contributions of the cluster
expansion are taken into account. The proximity to the experimental point
for, e.g. $I_{\vartheta}=0.09\,\text{fm}^{3}$, thus indicates, that the
omitted three-body terms of the cluster expansion can be tuned such that
they  cancel the contributions of the genuine three-body force to a large
extent. The influence of the genuine three-body force on the binding
energies in these small systems can therefore be minimized by a proper
choice of the correlators, i.e., by choosing the phase-shift equivalent
two-body force which needs the weakest three-body force. 
This, however, does not mean that the three-body force can be avoided
completely. One should keep in mind that the above observations 
refer to a single observable and to very small systems only.

\section{Conclusions}

The Unitary Correlation Operator Method provides a powerful and
transparent tool to construct phase-shift equivalent low-momentum
interactions by means of an explicit unitary transformation of a realistic
NN-potential. The physics of short-range central and tensor correlations
is encapsulated in the optimal correlators which are determined in the
two-body system. For the long-ranged tensor component a separation of
short-range state-independent correlations and long-range correlations is
performed through an additional constraint on the range of the
correlators. Once the correlators are fixed, we can evaluate the unitary
transformation of either states or operators directly.

In the case of two-body matrix elements of the correlated Hamiltonian in an
$LS$-coupled basis, it is convenient to map the unitary correlators onto
the two-body angular momentum eigenstates. The resulting correlated matrix
elements reveal some of the important features of the correlated
interaction. The unitary transformation causes a pre-diagonalization of the
Hamiltonian, i.e., large off-diagonal momentum-space matrix elements induced 
by the central core and the tensor interaction are eliminated and
non-vanishing matrix elements remain solely in a band along the diagonal.

Correlated matrix elements, e.g. with respect to a harmonic oscillator basis,
serve as universal input for different many-body calculations  
\footnote{An optimized code for the calculation of the correlated
oscillator matrix elements is available from the authors upon request.}. 
We have demonstrated the use of those matrix elements in the no-core shell
model for $A\leq4$. In comparison to a shell-model diagonalization with
the bare  interaction we observe a dramatic reduction of the ground state
energy in very small model-spaces and a significant improvement of
convergence with increasing size of the model space. Here the interplay
between unitary correlator and model-space becomes evident: The unitary
correlation operator describes the state-independent short-range
correlations induced by the central and the tensor part of the interaction
--- this accounts for the bulk of the binding energy.  State-dependent
long-range correlations are described by the model-space --- this leads to
the moderate gain in binding energy with increasing model space size. The
rapid convergence indicates that those long-range correlations, unlike the
short-range correlations, can be quite easily treated in small and
computationally accessible model-spaces.

The no-core shell model calculations also provide a guideline for the
choice of the range-constraint for the tensor correlators. As function of
the range of the tensor correlators a manifold of phase-shift equivalent
potentials is generated which map out the Tjon line. The correlator range
can be chosen such that the exact ground state energies for $A\leq4$ are
in good agreement with experiment. Thus the net impact of the residual
three-nucleon force on the binding energies in these small systems can be
minimized. 

The next step is to  use the correlated matrix elements as input for
nuclear structure calculations also for heavier isotopes. In a forthcoming
publication we will discuss Hartree-Fock as well as RPA calculations
across the whole nuclear chart using the same correlated realistic
interactions.

\section*{Acknowledgments}

We are very grateful to Petr Navr\'atil for providing us with the no-core
shell model code used in Sec. \ref{sec:ncsm}. We thank Achim Schwenk for
the $V_{\text{low}k}$ matrix elements used in Fig. \ref{fig:momentumme_vlowk}.
This work is supported by the Deutsche Forschungsgemeinschaft through
contract SFB 634.



\begin{thebibliography}{30}
\expandafter\ifx\csname natexlab\endcsname\relax\def\natexlab#1{#1}\fi
\expandafter\ifx\csname bibnamefont\endcsname\relax
  \def\bibnamefont#1{#1}\fi
\expandafter\ifx\csname bibfnamefont\endcsname\relax
  \def\bibfnamefont#1{#1}\fi
\expandafter\ifx\csname citenamefont\endcsname\relax
  \def\citenamefont#1{#1}\fi
\expandafter\ifx\csname url\endcsname\relax
  \def\url#1{\texttt{#1}}\fi
\expandafter\ifx\csname urlprefix\endcsname\relax\def\urlprefix{URL }\fi
\providecommand{\bibinfo}[2]{#2}
\providecommand{\eprint}[2][]{\url{#2}}

\bibitem[{\citenamefont{Wiringa et~al.}(1995)\citenamefont{Wiringa, Stoks, and
  Schiavilla}}]{WiSt95}
\bibinfo{author}{\bibfnamefont{R.~B.} \bibnamefont{Wiringa}},
  \bibinfo{author}{\bibfnamefont{V.~G.~J.} \bibnamefont{Stoks}},
  \bibnamefont{and}
  \bibinfo{author}{\bibfnamefont{R.}~\bibnamefont{Schiavilla}},
  \bibinfo{journal}{Phys. Rev. C} \textbf{\bibinfo{volume}{51}},
  \bibinfo{pages}{38} (\bibinfo{year}{1995}).

\bibitem[{\citenamefont{Machleidt}(2001)}]{Mach01}
\bibinfo{author}{\bibfnamefont{R.}~\bibnamefont{Machleidt}},
  \bibinfo{journal}{Phys. Rev. C} \textbf{\bibinfo{volume}{63}},
  \bibinfo{pages}{024001} (\bibinfo{year}{2001}).

\bibitem[{\citenamefont{Stoks et~al.}(1994)\citenamefont{Stoks, Klomp,
  Terheggen, and de~Swart}}]{StKl94}
\bibinfo{author}{\bibfnamefont{V.~G.~J.} \bibnamefont{Stoks}},
  \bibinfo{author}{\bibfnamefont{R.~A.~M.} \bibnamefont{Klomp}},
  \bibinfo{author}{\bibfnamefont{C.~P.~F.} \bibnamefont{Terheggen}},
  \bibnamefont{and} \bibinfo{author}{\bibfnamefont{J.~J.}
  \bibnamefont{de~Swart}}, \bibinfo{journal}{Phys. Rev. C}
  \textbf{\bibinfo{volume}{49}}, \bibinfo{pages}{2950} (\bibinfo{year}{1994}).

\bibitem[{\citenamefont{Pieper et~al.}(2002)\citenamefont{Pieper, Varga, and
  Wiringa}}]{PiVa02}
\bibinfo{author}{\bibfnamefont{S.~C.} \bibnamefont{Pieper}},
  \bibinfo{author}{\bibfnamefont{K.}~\bibnamefont{Varga}}, \bibnamefont{and}
  \bibinfo{author}{\bibfnamefont{R.~B.} \bibnamefont{Wiringa}},
  \bibinfo{journal}{Phys. Rev. C} \textbf{\bibinfo{volume}{66}},
  \bibinfo{pages}{044310} (\bibinfo{year}{2002}).

\bibitem[{\citenamefont{Pieper and Wiringa}(2001)}]{PiWi01}
\bibinfo{author}{\bibfnamefont{S.~C.} \bibnamefont{Pieper}} \bibnamefont{and}
  \bibinfo{author}{\bibfnamefont{R.~B.} \bibnamefont{Wiringa}},
  \bibinfo{journal}{Ann. Rev. Nucl. Part. Sci.} \textbf{\bibinfo{volume}{51}},
  \bibinfo{pages}{53} (\bibinfo{year}{2001}).

\bibitem[{\citenamefont{Pieper et~al.}(2004)\citenamefont{Pieper, Wiringa, and
  Carlson}}]{PiWi04}
\bibinfo{author}{\bibfnamefont{S.~C.} \bibnamefont{Pieper}},
  \bibinfo{author}{\bibfnamefont{R.~B.} \bibnamefont{Wiringa}},
  \bibnamefont{and} \bibinfo{author}{\bibfnamefont{J.}~\bibnamefont{Carlson}},
  \bibinfo{journal}{Phys. Rev. C} \textbf{\bibinfo{volume}{70}},
  \bibinfo{pages}{054325} (\bibinfo{year}{2004}).

\bibitem[{\citenamefont{Caurier et~al.}(2002)\citenamefont{Caurier, Navr\'atil,
  Ormand, and Vary}}]{CaNa02}
\bibinfo{author}{\bibfnamefont{E.}~\bibnamefont{Caurier}},
  \bibinfo{author}{\bibfnamefont{P.}~\bibnamefont{Navr\'atil}},
  \bibinfo{author}{\bibfnamefont{W.~E.} \bibnamefont{Ormand}},
  \bibnamefont{and} \bibinfo{author}{\bibfnamefont{J.~P.} \bibnamefont{Vary}},
  \bibinfo{journal}{Phys. Rev. C} \textbf{\bibinfo{volume}{66}},
  \bibinfo{pages}{024314} (\bibinfo{year}{2002}).

\bibitem[{\citenamefont{Navr\'atil and Ormand}(2002)}]{NaOr02}
\bibinfo{author}{\bibfnamefont{P.}~\bibnamefont{Navr\'atil}} \bibnamefont{and}
  \bibinfo{author}{\bibfnamefont{W.~E.} \bibnamefont{Ormand}},
  \bibinfo{journal}{Phys. Rev. Lett.} \textbf{\bibinfo{volume}{88}},
  \bibinfo{pages}{152502} (\bibinfo{year}{2002}).

\bibitem[{\citenamefont{Navr\'atil
  et~al.}(2000{\natexlab{a}})\citenamefont{Navr\'atil, Vary, and
  Barrett}}]{NaVa00b}
\bibinfo{author}{\bibfnamefont{P.}~\bibnamefont{Navr\'atil}},
  \bibinfo{author}{\bibfnamefont{J.~P.} \bibnamefont{Vary}}, \bibnamefont{and}
  \bibinfo{author}{\bibfnamefont{B.~R.} \bibnamefont{Barrett}},
  \bibinfo{journal}{Phys. Rev. C} \textbf{\bibinfo{volume}{62}},
  \bibinfo{pages}{054311} (\bibinfo{year}{2000}{\natexlab{a}}).

\bibitem[{\citenamefont{Entem and Machleidt}(2003)}]{EnMa03}
\bibinfo{author}{\bibfnamefont{D.~R.} \bibnamefont{Entem}} \bibnamefont{and}
  \bibinfo{author}{\bibfnamefont{R.}~\bibnamefont{Machleidt}},
  \bibinfo{journal}{Phys. Rev C.} \textbf{\bibinfo{volume}{68}},
  \bibinfo{pages}{041001(R)} (\bibinfo{year}{2003}).

\bibitem[{\citenamefont{Epelbaum et~al.}(2002)\citenamefont{Epelbaum, Nogga,
  Gl\"ockle, Kamada, Mei\ss{}ner, and Witala}}]{EpNo02}
\bibinfo{author}{\bibfnamefont{E.}~\bibnamefont{Epelbaum}},
  \bibinfo{author}{\bibfnamefont{A.}~\bibnamefont{Nogga}},
  \bibinfo{author}{\bibfnamefont{W.}~\bibnamefont{Gl\"ockle}},
  \bibinfo{author}{\bibfnamefont{H.}~\bibnamefont{Kamada}},
  \bibinfo{author}{\bibfnamefont{Ulf-G.} \bibnamefont{Mei\ss{}ner}},
  \bibnamefont{and} \bibinfo{author}{\bibfnamefont{H.}~\bibnamefont{Witala}},
  \bibinfo{journal}{Phys. Rev. C} \textbf{\bibinfo{volume}{66}},
  \bibinfo{pages}{064001} (\bibinfo{year}{2002}).

\bibitem[{\citenamefont{Bogner et~al.}(2003{\natexlab{a}})\citenamefont{Bogner,
  Kuo, and Schwenk}}]{BoKu03}
\bibinfo{author}{\bibfnamefont{S.~K.} \bibnamefont{Bogner}},
  \bibinfo{author}{\bibfnamefont{T.~T.~S.} \bibnamefont{Kuo}},
  \bibnamefont{and} \bibinfo{author}{\bibfnamefont{A.}~\bibnamefont{Schwenk}},
  \bibinfo{journal}{Phys. Rep.} \textbf{\bibinfo{volume}{386}},
  \bibinfo{pages}{1} (\bibinfo{year}{2003}{\natexlab{a}}).

\bibitem[{\citenamefont{Bogner et~al.}(2003{\natexlab{b}})\citenamefont{Bogner,
  Kuo, Schwenk, Entem, and Machleidt}}]{BoKu03b}
\bibinfo{author}{\bibfnamefont{S.~K.} \bibnamefont{Bogner}},
  \bibinfo{author}{\bibfnamefont{T.~T.~S.} \bibnamefont{Kuo}},
  \bibinfo{author}{\bibfnamefont{A.}~\bibnamefont{Schwenk}},
  \bibinfo{author}{\bibfnamefont{D.~R.} \bibnamefont{Entem}}, \bibnamefont{and}
  \bibinfo{author}{\bibfnamefont{R.}~\bibnamefont{Machleidt}},
  \bibinfo{journal}{Phys. Lett. B} \textbf{\bibinfo{volume}{576}},
  \bibinfo{pages}{265} (\bibinfo{year}{2003}{\natexlab{b}}).

\bibitem[{\citenamefont{Feldmeier et~al.}(1998)\citenamefont{Feldmeier, Neff,
  Roth, and Schnack}}]{FeNe98}
\bibinfo{author}{\bibfnamefont{H.}~\bibnamefont{Feldmeier}},
  \bibinfo{author}{\bibfnamefont{T.}~\bibnamefont{Neff}},
  \bibinfo{author}{\bibfnamefont{R.}~\bibnamefont{Roth}}, \bibnamefont{and}
  \bibinfo{author}{\bibfnamefont{J.}~\bibnamefont{Schnack}},
  \bibinfo{journal}{Nucl. Phys.} \textbf{\bibinfo{volume}{A632}},
  \bibinfo{pages}{61} (\bibinfo{year}{1998}).

\bibitem[{\citenamefont{Neff and Feldmeier}(2003)}]{NeFe03}
\bibinfo{author}{\bibfnamefont{T.}~\bibnamefont{Neff}} \bibnamefont{and}
  \bibinfo{author}{\bibfnamefont{H.}~\bibnamefont{Feldmeier}},
  \bibinfo{journal}{Nucl. Phys.} \textbf{\bibinfo{volume}{A713}},
  \bibinfo{pages}{311} (\bibinfo{year}{2003}).

\bibitem[{\citenamefont{Roth et~al.}(2004)\citenamefont{Roth, Neff, Hergert,
  and Feldmeier}}]{RoNe04}
\bibinfo{author}{\bibfnamefont{R.}~\bibnamefont{Roth}},
  \bibinfo{author}{\bibfnamefont{T.}~\bibnamefont{Neff}},
  \bibinfo{author}{\bibfnamefont{H.}~\bibnamefont{Hergert}}, \bibnamefont{and}
  \bibinfo{author}{\bibfnamefont{H.}~\bibnamefont{Feldmeier}},
  \bibinfo{journal}{Nucl. Phys. A} \textbf{\bibinfo{volume}{745}},
  \bibinfo{pages}{3} (\bibinfo{year}{2004}).

\bibitem[{\citenamefont{Neff and Feldmeier}(2004)}]{NeFe04}
\bibinfo{author}{\bibfnamefont{T.}~\bibnamefont{Neff}} \bibnamefont{and}
  \bibinfo{author}{\bibfnamefont{H.}~\bibnamefont{Feldmeier}},
  \bibinfo{journal}{Nucl. Phys. A} \textbf{\bibinfo{volume}{738}},
  \bibinfo{pages}{357} (\bibinfo{year}{2004}).

\bibitem[{\citenamefont{Neff et~al.}(2005)\citenamefont{Neff, Feldmeier, and
  Roth}}]{NeFe05}
\bibinfo{author}{\bibfnamefont{T.}~\bibnamefont{Neff}},
  \bibinfo{author}{\bibfnamefont{H.}~\bibnamefont{Feldmeier}},
  \bibnamefont{and} \bibinfo{author}{\bibfnamefont{R.}~\bibnamefont{Roth}},
  \bibinfo{journal}{Nucl. Phys. A} \textbf{\bibinfo{volume}{752}},
  \bibinfo{pages}{321} (\bibinfo{year}{2005}).

\bibitem[{\citenamefont{Suzuki and Lee}(1980)}]{SuLe80}
\bibinfo{author}{\bibfnamefont{K.}~\bibnamefont{Suzuki}} \bibnamefont{and}
  \bibinfo{author}{\bibfnamefont{S.~Y.} \bibnamefont{Lee}},
  \bibinfo{journal}{Prog. Theo. Phys.} \textbf{\bibinfo{volume}{64}},
  \bibinfo{pages}{2091} (\bibinfo{year}{1980}).

\bibitem[{\citenamefont{Fujii et~al.}(2004)\citenamefont{Fujii, Okamoto, and
  Suzuki}}]{FuOk04}
\bibinfo{author}{\bibfnamefont{S.}~\bibnamefont{Fujii}},
  \bibinfo{author}{\bibfnamefont{R.}~\bibnamefont{Okamoto}}, \bibnamefont{and}
  \bibinfo{author}{\bibfnamefont{K.}~\bibnamefont{Suzuki}},
  \bibinfo{journal}{Phys. Rev. C} \textbf{\bibinfo{volume}{69}},
  \bibinfo{pages}{034328} (\bibinfo{year}{2004}).

\bibitem[{\citenamefont{Roth}(2000)}]{Roth00}
\bibinfo{author}{\bibfnamefont{R.}~\bibnamefont{Roth}}, Ph.D. thesis,
  \bibinfo{school}{Technische Universit\"at Darmstadt} (\bibinfo{year}{2000}).

\bibitem[{\citenamefont{Machleidt}(1989)}]{Mach89}
\bibinfo{author}{\bibfnamefont{R.}~\bibnamefont{Machleidt}},
  \bibinfo{journal}{Adv. Nucl. Phys.} \textbf{\bibinfo{volume}{19}},
  \bibinfo{pages}{189} (\bibinfo{year}{1989}).

\bibitem[{\citenamefont{Feldmeier}(1990)}]{Feld90}
\bibinfo{author}{\bibfnamefont{H.}~\bibnamefont{Feldmeier}},
  \bibinfo{journal}{Nucl. Phys.} \textbf{\bibinfo{volume}{A515}},
  \bibinfo{pages}{147} (\bibinfo{year}{1990}).

\bibitem[{\citenamefont{Feldmeier and Schnack}(2000)}]{FeSc00}
\bibinfo{author}{\bibfnamefont{H.}~\bibnamefont{Feldmeier}} \bibnamefont{and}
  \bibinfo{author}{\bibfnamefont{J.}~\bibnamefont{Schnack}},
  \bibinfo{journal}{Rev. Mod. Phys.} \textbf{\bibinfo{volume}{72}},
  \bibinfo{pages}{655} (\bibinfo{year}{2000}).

\bibitem[{\citenamefont{Stetcu et~al.}(2005)\citenamefont{Stetcu, Barrett,
  Navr\'atil, and Vary}}]{StBa05}
\bibinfo{author}{\bibfnamefont{I.}~\bibnamefont{Stetcu}},
  \bibinfo{author}{\bibfnamefont{B.~R.} \bibnamefont{Barrett}},
  \bibinfo{author}{\bibfnamefont{P.}~\bibnamefont{Navr\'atil}},
  \bibnamefont{and} \bibinfo{author}{\bibfnamefont{J.~P.} \bibnamefont{Vary}},
  \bibinfo{journal}{Phys. Rev. C} \textbf{\bibinfo{volume}{71}},
  \bibinfo{pages}{044325} (\bibinfo{year}{2005}).

\bibitem[{\citenamefont{Machleidt and Slaus}(2001)}]{MaSl01}
\bibinfo{author}{\bibfnamefont{R.}~\bibnamefont{Machleidt}} \bibnamefont{and}
  \bibinfo{author}{\bibfnamefont{I.}~\bibnamefont{Slaus}}, \bibinfo{journal}{J.
  Phys. G: Nucl. Part. Phys.} \textbf{\bibinfo{volume}{27}},
  \bibinfo{pages}{R69} (\bibinfo{year}{2001}).

\bibitem[{\citenamefont{Vary et~al.}(1973)\citenamefont{Vary, Sauer, and
  Wong}}]{VaSa73}
\bibinfo{author}{\bibfnamefont{J.~P.} \bibnamefont{Vary}},
  \bibinfo{author}{\bibfnamefont{P.~U.} \bibnamefont{Sauer}}, \bibnamefont{and}
  \bibinfo{author}{\bibfnamefont{C.~W.} \bibnamefont{Wong}},
  \bibinfo{journal}{Phys. Rev. C} \textbf{\bibinfo{volume}{7}},
  \bibinfo{pages}{1776} (\bibinfo{year}{1973}).

\bibitem[{\citenamefont{Navr\'atil
  et~al.}(2000{\natexlab{b}})\citenamefont{Navr\'atil, Kamuntavicius, and
  Barrett}}]{NaKa00}
\bibinfo{author}{\bibfnamefont{P.}~\bibnamefont{Navr\'atil}},
  \bibinfo{author}{\bibfnamefont{G.~P.} \bibnamefont{Kamuntavicius}},
  \bibnamefont{and} \bibinfo{author}{\bibfnamefont{B.~R.}
  \bibnamefont{Barrett}}, \bibinfo{journal}{Phys. Rev. C}
  \textbf{\bibinfo{volume}{61}}, \bibinfo{pages}{044001}
  (\bibinfo{year}{2000}{\natexlab{b}}).

\bibitem[{\citenamefont{Nogga et~al.}(2000)\citenamefont{Nogga, Kamada, and
  Gl"ockle}}]{NoKa00}
\bibinfo{author}{\bibfnamefont{A.}~\bibnamefont{Nogga}},
  \bibinfo{author}{\bibfnamefont{H.}~\bibnamefont{Kamada}}, \bibnamefont{and}
  \bibinfo{author}{\bibfnamefont{W.}~\bibnamefont{Gl\"ockle}},
  \bibinfo{journal}{Phys. Rev. Lett.} \textbf{\bibinfo{volume}{85}},
  \bibinfo{pages}{944} (\bibinfo{year}{2000}).

\bibitem[{\citenamefont{Nogga et~al.}(2004)\citenamefont{Nogga, Bogner, and
  Schwenk}}]{NoBo04}
\bibinfo{author}{\bibfnamefont{A.}~\bibnamefont{Nogga}},
  \bibinfo{author}{\bibfnamefont{S.~K.} \bibnamefont{Bogner}},
  \bibnamefont{and} \bibinfo{author}{\bibfnamefont{A.}~\bibnamefont{Schwenk}},
  \bibinfo{journal}{Phys. Rev. C} \textbf{\bibinfo{volume}{70}},
  \bibinfo{pages}{061002(R)} (\bibinfo{year}{2004}).

\end{thebibliography}
\end{document}